\documentclass[journal=chreay, manuscript=review, layout=twocolumn]{achemso} 
\usepackage{cuted}
\usepackage{amsmath}
\usepackage{amssymb}
\usepackage{cases}
\usepackage{enumitem}
\usepackage{color}
\usepackage[toc,page]{appendix}
\usepackage{graphicx,subfigure}
\usepackage{float}
\usepackage{bm}
\usepackage{array}

\usepackage{amsmath}
\usepackage{cuted}    
\usepackage{etoolbox} 
\usepackage{flushend} 
\usepackage{balance}  

\makeatletter
\patchcmd{\@outputdblcol}
  {\global \@twocolumnfalse}
  {\global\@twocolumnfalse\close@column@grid}
  {}{}

\newenvironment{widetext}{
  \begin{strip}
    \rule{\hsize}{0.4pt}\vspace{0.5em}
  }{
    \vspace{0.5em}\rule{\hsize}{0.4pt}
  \end{strip}
}
\makeatother

\usepackage{hyperref}
\hypersetup{
    colorlinks=false, 
    linkcolor=cyan,
    citecolor=magenta, 
    filecolor=magenta, 
    urlcolor=cyan,
    runcolor=cyan
}
\usepackage[capitalise]{cleveref}
\usepackage{titlesec}
\titleclass{\subsubsubsection}{straight}[\subsubsection]

\newcounter{subsubsubsection}[subsubsection]
\renewcommand\thesubsubsubsection{\thesubsubsection.\arabic{subsubsubsection}}

\titleformat{\paragraph}
  {\normalfont\normalsize\itshape}{\thesubsubsubsection}{1em}{}

\titlespacing*{\paragraph}
{0pt}{1.25ex plus 1ex minus .2ex}{0.5ex plus .1ex}

\newcommand{\ignore}[1]{}

\def\mc#1{\mathcal{#1}}
\def\bra#1{\langle{#1}|}
\def\ket#1{|{#1}\rangle}
\def\superbra#1{\langle\langle{#1}|}
\def\superket#1{|{#1}\rangle\rangle}
\def\braket#1{\langle{#1}\rangle}

\def\>{\rangle}
\def\<{\langle}
\def\Tr{\operatorname{Tr}}
\def\mcBH{\mathcal{B}(\mathcal{H})}
\newcommand{\ketb}[2]{|{#1}\>\!\<#2|}

\newcommand{\cU}{\mathcal{U}}

\newcommand{\tP}{\tilde{\Pi}}

\newcommand{\rP}{\rho_{\Pi}}
\newcommand{\rtP}{\rho_{\tP}}
\newcommand{\sk}{\sigma_k}
\newcommand{\tsk}{\tilde{\sk}}

\usepackage{titlesec}
\usepackage{multicol}

\newtheorem{mytheorem}{Theorem}

\newcommand{\bes} {\begin{subequations}}
\newcommand{\ees} {\end{subequations}}
\newcommand{\ba}{\begin{eqnarray}}
\newcommand{\ea}{\end{eqnarray}}
\newcommand{\bea} {\begin{eqnarray}}
\newcommand{\eea} {\end{eqnarray}}
\newcommand{\beq}{\begin{equation}}
\newcommand{\eeq}{\end{equation}}

\newcommand{\eps} {\varepsilon}
\newcommand{\epstot} {\eps_{\mathrm{tot}}}
\newcommand{\epserr} {\eps_{\mathrm{err}}}
\newcommand{\delerr} {\Delta_{\mathrm{err}}}
\newcommand{\therr}{\theta_{\mathrm{err}}}
\newcommand{\dphi}{\delta\phi}
\newcommand{\dth}{\delta\theta}
\newcommand{\bX}{\overline{X}}
\newcommand{\bY}{\overline{Y}}
\newcommand{\bv}{\mathbf{v}}
\newcommand{\bh}{\mathbf{h}}
\newcommand{\bc}{\mathbf{c}}
\newcommand{\bz}{\boldsymbol\zeta}
\newcommand{\bx}{\boldsymbol\xi}
\newcommand{\stkout}[1]{\ifmmode\text{\sout{\ensuremath{#1}}}\else\sout{#1}\fi}

\newcolumntype{P}[1]{>{\centering\arraybackslash}p{#1}}

\usepackage{appendix}
\usepackage{commath}
\usepackage{color}
\usepackage{amsmath}
\usepackage{adjustbox}
\usepackage[normalem]{ulem}
\usepackage{color}
\usepackage{threeparttable}
\usepackage{booktabs}
\usepackage{array}
\usepackage{siunitx}
\usepackage{makecell}
\usepackage[dvipsnames]{xcolor}
\usepackage[normalem]{ulem}

\author{Vinay Tripathi}
\affiliation{Department of Physics \& Astronomy, University of Southern California, Los Angeles, California 90089, USA}
\alsoaffiliation{Center for Quantum Information Science and Technology, University of Southern California, Los Angeles, California 90089, USA}
\altaffiliation{These two authors contributed equally to this work.}

\author{Daria Kowsari}

\affiliation{Department of Physics \& Astronomy, University of Southern California, Los Angeles, California 90089, USA}
\alsoaffiliation{Center for Quantum Information Science and Technology, University of Southern California, Los Angeles, California 90089, USA}
\altaffiliation{These two authors contributed equally to this work.}

\author{Kumar Saurav}
\affiliation{Center for Quantum Information Science and Technology, University of Southern California, Los Angeles, California 90089, USA}
\alsoaffiliation{Ming Hsieh Department of Electrical \& Computer Engineering, University of Southern California, Los Angeles, California 90089, USA}

\author{Haimeng Zhang}
\affiliation{Center for Quantum Information Science and Technology, University of Southern California, Los Angeles, California 90089, USA}
\alsoaffiliation{Ming Hsieh Department of Electrical \& Computer Engineering, University of Southern California, Los Angeles, California 90089, USA}

\author{\\Eli M. Levenson-Falk}
\email{elevenso@usc.edu}
\affiliation{Department of Physics \& Astronomy, University of Southern California, Los Angeles, California 90089, USA}
\alsoaffiliation{Center for Quantum Information Science and Technology, University of Southern California, Los Angeles, California 90089, USA}
\alsoaffiliation{Ming Hsieh Department of Electrical \& Computer Engineering, University of Southern California, Los Angeles, California 90089, USA}

\author{Daniel A. Lidar}
\email{lidar@usc.edu}
\affiliation{Department of Physics \& Astronomy, University of Southern California, Los Angeles, California 90089, USA}
\alsoaffiliation{Center for Quantum Information Science and Technology, University of Southern California, Los Angeles, California 90089, USA}
\alsoaffiliation{Ming Hsieh Department of Electrical \& Computer Engineering, University of Southern California, Los Angeles, California 90089, USA}
\alsoaffiliation{Department of Chemistry, University of Southern California, Los Angeles, California 90089, USA}

\title{Benchmarking quantum gates and circuits}

\begin{document}

\begin{abstract}
Accurate noise characterization in quantum gates and circuits is vital for the development of reliable quantum simulations for chemically relevant systems and fault-tolerant quantum computing. This paper reviews a variety of key benchmarking techniques, including Randomized Benchmarking, Quantum Process Tomography, Gate Set Tomography, Process Fidelity Estimation, Direct Fidelity Estimation, and Cross-Entropy Benchmarking. We evaluate each method's complexities, the resources they require, and their effectiveness in addressing coherent, incoherent, and state preparation and measurement (SPAM) errors. Furthermore, we introduce \textit{Deterministic Benchmarking} (DB), a novel protocol that minimizes the number of experimental runs, exhibits resilience to SPAM errors, and effectively characterizes both coherent and incoherent errors. The implementation of DB is experimentally validated using a superconducting transmon qubit, and the results are substantiated with a simple analytical model and master equation simulations. With the addition
of DB to the toolkit of available benchmarking methods, this article serves as a practical guide for choosing and applying benchmarking protocols to advance quantum computing technologies.
\end{abstract}

\tableofcontents

\section{Introduction}

The realization of the promise of quantum computation~\cite{Bharti:2022aa,dalzell2023quantumalgorithmssurveyapplications} 
relies on the development of fault-tolerant, error-corrected quantum computers. Fault-tolerance is essential for the ultimate scalability of quantum computing, but requires high-fidelity operations that exceed an accuracy threshold \cite{Shor1996, Aharonov:96, Knill:98, Aliferis:05, Campbell:2017aa}. 
When targeting chemical accuracy (1kcal/mol) for electronic structure simulations, the impact of gate error becomes magnified because small inaccuracies in energies ---  especially for strongly correlated molecules --- can alter reaction pathways and observables. Error probabilities per gate below $10^{–4}$ could be vital in large-scale fault-tolerant simulations of industrially relevant catalysts, as higher error rates dramatically inflate the number of logical qubits and error-correcting cycles required to achieve a final result with the desired precision~\cite{Reiher:2017ab}. Similarly, in quantum dynamics calculations, which often involve iterated time-step evolution, the accumulated error can become prohibitively large unless physical error rates are suppressed to the $10^{–5}-10^{–6}$ regime~\cite{annurev-physchem-032210-103512}. In these scenarios, each additional order of magnitude in gate fidelity can translate into disproportionately lower qubit counts or reduced circuit depth, thereby expanding the range of molecular systems amenable to quantum simulation.

Characterizing quantum operations is thus crucial and plays a pivotal role in the development and implementation of universal quantum computers. Accurate characterization allows for the identification and subsequent improvement of errors in quantum gates. 
But even before the advent of universal, fault tolerant quantum computation, today's noisy intermediate scale quantum (NISQ) computers~\cite{Preskill2018} are being actively investigated in terms of quantum chemistry applications~\cite{OMalley:2016aa,Kandala:2017aa}. Early algorithms focused on exact diagonalization~\cite{Lidar:98RC,Aspuru-Guzik:05}, while much current attention is devoted to variational algorithms~\cite{Peruzzo:2014aa,McClean:2016aa}. From a control-theoretic perspective, reducing gate errors --- particularly multi-qubit errors that can amplify coherent leakage or correlated noise channels --- is indispensable for accurate electronic structure calculations~\cite{McArdle:2020aa}. Near-term algorithms such as the Variational Quantum Eigensolver (VQE)~\cite{Tilly:2022aa} and, eventually, more resource-intensive protocols like Quantum Phase Estimation (QPE)~\cite{kitaev1995quantummeasurementsabelianstabilizer}, hinge on efficient gate decompositions that preserve delicate coherence over increasingly large circuit depths~\cite{Cao:2019aa}. Any deviation in gate fidelity can accumulate into significant numerical biases, thus compromising predictions of reaction energies or catalytic mechanisms~\cite{Lee:2023aa}. Consequently, noise-reduction strategies have emerged as key tools in the quest to push current hardware toward the threshold of chemical accuracy~\cite{Huggins:2021aa}. For such techniques to function and improve performance, accurate characterization and benchmarking of quantum operations are essential.

Several methods have been developed over the years for benchmarking quantum operations, particularly focusing on quantum gates. These methods are part of the broader research field of Quantum Characterization, Verification and Validation (QCVV) \cite{hashim2024}, and include Quantum Process Tomography (QPT) \cite{chuang-sqpt,poyatos-sqpt,Mohseni:2008ly}, Randomized Benchmarking (RB) and its many variants \cite{Emerson_2005, Knill2008, Magesan2011, Helsen:2022aa}, Gate Set Tomography (GST) \cite{ibm_gst_2013, sandia_gst_2013, sandia_gst_2014}, Process Fidelity Estimation~\cite{baumer2024quantum}, Direct Fidelity Estimation~\cite{da_Silva_Landon_Cardinal_Poulin_2011,Flammia_Liu_2011}, and Cross-Entropy Benchmarking~\cite{neill_blueprint_2018,boixo_characterizing_2018,arute_quantum_2019}.  
QPT provides a full reconstruction of a quantum process, but is resource-intensive and scales poorly with system size. RB, on the other hand, offers a scalable alternative by estimating the average error rate of quantum gates, but it lacks the ability to identify specific error types. GST addresses some of these limitations by providing a self-consistent characterization of a complete set of quantum gates, albeit at the cost of increased experimental complexity. 

Gate errors are fundamentally either coherent or incoherent. Coherent errors that arise from systematic deviations in quantum gate operations pose a significant challenge in the context of error correction\cite{Kueng2016, Barnes2017}. While stochastic errors add up as probabilities, coherent errors add up as amplitudes, generally leading to a quadratically smaller accuracy threshold \cite{ng:032318}. Traditional error characterization methods such as RB, which deliberately randomize noise maps, do not capture coherent error accumulation unless appropriately modified \cite{Wallman:2015aa}. This can result in an underestimation of the gate performance needed to achieve accuracy thresholds, critical for achieving fault-tolerant quantum computing \cite{Sanders2016, Hashim2023}. Various techniques exist that can mitigate the accumulation of coherent errors by converting them into stochastic noise, but only at the expense of significant additional cost in the number of required circuits \cite{Wallman2016, Hashim2021, Berg2023, seif2403.06852}. It can be advantageous to, instead,  characterize coherent gate errors~\cite{Gross2024}, and subsequently eliminate or at least suppress such errors at the control level, rather than placing the burden of doing so on fault-tolerant quantum error correction using large codes \cite{Bravyi2018}.

In this Review, we provide a survey of quantum gate characterization methods, emphasizing the role of each strategy depending on the end goal. As the field is vast, we sacrifice generality somewhat by emphasizing the characterization of single-qubit and two-qubit gates, given that these are the fundamental building blocks of quantum circuits\cite{Barenco:1995aa}. However, we also include methods designed to provide holistic benchmarks at the algorithm level. 

After providing, in \cref{sec:gate-errors}, a general introduction to the sources of gate errors and the formalism for characterizing them, we cover in \cref{sec:protocols}, a variety of methods ranging from the most comprehensive and resource-intensive, such as quantum process tomography and gate set tomography, to the least detailed, such as randomized benchmarking. We discuss the trade-offs between these techniques, their scalability, and their applicability to various quantum computing architectures. 

To keep this review focused, we refrain from providing a comprehensive analysis of all available benchmarking methods. Consequently, we do not discuss several significant methods, such as cycle benchmarking\cite{Erhard_Wallman_Postler_Meth_Stricker_Martinez_Schindler_Monz_Emerson_Blatt_2019}, leakage and crosstalk characterization\cite{PhysRevA.96.022330}, quantum volume benchmarking\cite{PhysRevA.100.032328}, fidelity estimation with 2-designs\cite{PhysRevA.80.012304}, mirror benchmarking\cite{Proctor_2021}, variational quantum benchmarking\cite{Cincio_2018}, average gate fidelity\cite{PhysRevLett.111.200401}, machine learning-based benchmarking\cite{Torlai_2018,Koutn__2022}, channel spectrum benchmarking\cite{Gu:2023aa}, and error per layered gate (EPLG) \cite{mckay2023benchmarking}, among others. Most of these methods are variants of the key methods we focus on here, so we hope that the reader will feel well-prepared to tackle these other methods after reading this review.

However, this work is more than a review, as we also introduce an original benchmarking protocol that we call \emph{deterministic benchmarking} (DB). This new protocol, which is appropriate for characterizing single-qubit gates, requires fewer experimental runs than previous characterization methods, exhibits robustness to SPAM errors, and provides detailed information about coherent errors. We discuss DB in detail in \cref{sec:DB}, provide experimental evidence of its efficacy, and contrast it with RB.

\section{Gate errors}
\label{sec:gate-errors}

\subsection{Quantum dynamical maps}
\label{sec:QDM}

Assuming a factorized initial
system-bath state, i.e., a system which is not entangled with its bath, the dynamics of an open quantum system with a $D$-dimensional Hilbert space $\mathcal{H}$ can be described
by a completely positive trace-preserving (CPTP) linear map~\cite{Breuer:book}:
\begin{equation}
\mathcal{E}(\rho )=\sum_{i}A_{i}\rho A_{i}^{\dagger },
\label{eq:E}
\end{equation}
where $\rho\in \mathcal{B}(\mathcal{H})$ (the space of linear operators acting on $\mathcal{H}$) is the initial state of the system and satisfies $\Tr(\rho)=1$ as well as $\rho\ge 0$ (its eigenvalues are all non-negative). The Kraus operators~\cite{Kraus:83} $A_i$ satisfy $\sum_{i}A^{\dagger}_{i}A_{i}= I$ (the identity operator), which guarantees that $\Tr[\mathcal{E}(\rho )]=1$. Let $\{E_{m}\}_{m=0}^{D^2-1}$ be a set of fixed
Hermitian basis operators for $\mathcal{B}(\mathcal{H})$, which satisfy the
orthogonality condition
\begin{equation}
\Tr (E_{m}^{\dagger }E_{n})=D\delta _{mn}.
\end{equation}
For example, we may choose $E_{0}=I$ and $E_{m}$, $m=1,\ldots ,D^2-1$, to be
traceless Hermitian matrices. For a multi-qubit system, the $E_{m}$'s can be tensor products
of the identity operator and the Pauli matrices, which we denote interchangeably by $\{\sigma^x,\sigma^y,\sigma^z\}$ or $\{X,Y,Z\}$. We also write Pauli tensor products interchangeably as $\sigma^x\otimes\sigma^y$ or $XY$, etc.

The $A_{i}$ operators can be decomposed as $A_{i}=\sum_{m}a_{im}E_{m}$, and therefore we have
\begin{equation}
\mathcal{E}(\rho )=\sum_{m,n=0}^{D^2-1}\chi _{mn}E_{m}\rho E_{n}^{\dagger },
\label{eq:chi-from-E}
\end{equation}
where $\chi _{mn}=\sum_{i}a_{in}^{\ast }a_{im} = \chi_{nm}^*$. The ``process matrix''
$\bm{\chi}$ is not only Hermitian but also positive semi-definite (all its eigenvalues are non-negative). It captures all the information about the map $\mathcal{E}$ with
respect to the $\{E_{i}\}$ basis, i.e., a characterization of $\mathcal{E}$ is
equivalent to a determination of all the
matrix elements of $\bm{\chi}$, where the $E_{i}$ play the role of observables. Since $\bm{\chi}$ is Hermitian and the map $\mathcal{E}$ is trace-preserving, the corresponding process matrix $\bm{\chi}$ has at most $D^4-D^2$
independent real parameters [see \cref{eq:gst_comprel} below]. We will focus most of our attention on the $n$-qubit case, i.e., $D=2^{n}$.

The measurement formalism of quantum mechanics is conveniently formulated using the language of CPTP maps as well~\cite{nielsen2010quantum}. A set of generalized measurement operators $\{A_i\}$ is any set in $\mathcal{B}(\mathcal{H})$ that satisfies the constraint $\sum_{i}A^{\dagger}_iA_{i}= I$. When a system in the state $\rho$ is measured, the post-measurement state is $\rho'_i=A_i\rho A_i^\dagger/p_i$ with probability $p_i = \Tr(A_i\rho A_i^\dagger) = \Tr(M_i\rho)$, where the set $\{M_i\} = \{A_i^\dagger A_i\}$ is known as a positive operator valued measure (POVM).
If the measurement outcome is not known, the resulting state becomes $\sum_i A_i \rho A_i^\dagger$.

The standard case of projective von-Neumann measurements is the special case when the generalized measurement operators are projectors, i.e., are Hermitian and satisfy $A_iA_j = \delta_{ij}A_i$. When an observable (any Hermitian operator) $\Omega$ with spectral decomposition $\Omega = \sum_k \lambda_k P_k$ ($\lambda_k$ and $P_k$ are its eigenvalues and eigenprojectors, respectively) is measured, the post-measurement state is $\rho'_k= P_k \rho P_k/p_k$ with probability $p_k = \Tr(P_k \rho)$, and the measurement outcome is the eigenvalue $\lambda_k$.
The state resulting from ignorance of the measurement result is then $\sum_k P_k \rho P_k$, which generally corresponds to a mixed state, i.e., a state $\sigma$ whose purity $\Tr(\sigma^2)$ is less than $1$.

The Choi-Jamio\l kowski isomorphism\cite{Jamiolkowski:72,Choi:75} is a correspondence between maps acting on states in $\mcBH$ and states in $\mc{H}\otimes\mc{H}$. More concretely, the Choi state $\rho_{\mc{E}}$ of a quantum map $\mc{E}$ is
\beq
\rho_{\mc{E}} = (\mathbb{I} \otimes {\mc{E}})(\ketb{\phi}{\phi}),
\label{eq:Choi-state}
\eeq
where $\ket{\phi} = \frac{1}{\sqrt{D}}\sum_{i=1}^D \ket{i}\otimes\ket{i}$ is a maximally entangled state. 
The Uhlmann fidelity~\cite{Uhlmann} between two states $\rho,\sigma\in\mcBH$ is 
\beq
F(\sigma,\rho)= \|\sqrt{\rho}\sqrt{\sigma}\|_1 = \Tr{\left((\sigma^{1/2} \rho\sigma^{1/2})^{1/2}\right)} ,
\eeq
where the norm is the trace-norm: $\|A\|_1 = \Tr|A| = \Tr (\sqrt{A^\dagger A})$. The Uhlmann fidelity is bounded between $0$ and $1$ and is unitarily invariant. Note that some authors define it as the square of our quantity.

The \emph{process fidelity} $F$ between two quantum maps $\mc{E}$ and $\mc{E}'$, acting on the same quantum system, with respective Choi states $\rho_{\mc{E}}$ and $\rho_{\mc{E}'}$, is given by the Uhlmann fidelity between the Choi states:
\begin{equation} 
F_{\text{proc}}(\mc{E},\mc{E}') = F(\rho_{\mc{E}},\rho_{\mc{E}'}) = \|\sqrt{\rho_{\mc{E}}}\sqrt{\rho_{\mc{E}'}}\|_1 .
\label{eq:F-process}
\end{equation}
$F_{\text{proc}}$ quantifies the overlap between the two maps' actions on all possible input states. A process fidelity value close to $1$ indicates that the maps are very similar in their effect on quantum states. This metric is very useful for quantifying the similarity between a desired process, such as an ideal gate, and its actual, noisy implementation.


\subsection{Coherent and incoherent errors}
A quantum gate aims to produce some unitary operation $U$ which can be written as a superoperator $\mathcal{U}$ acting on the state $\rho$, i.e., $\mathcal{U}(\rho) = U\rho U^\dagger$. In practice, the actual operation is $\tilde{\mathcal{U}} = \mathcal{E} \circ \mathcal{U}$, where $\mathcal{E}$ now represents all errors on the gate. These errors can be divided into \emph{coherent} errors, where state purity is preserved but the unitary is altered, and \emph{incoherent} (stochastic) errors, where state purity [$\Tr(\rho^2)=1$] is lost due to some randomness in the evolution (i.e., decoherence). Put another way, coherent errors are deterministically repeatable, while incoherent errors give random evolution. Coherent errors are generally more damaging, as their amplitudes add in phase and can thus increase the gate failure probability quadratically faster than incoherent errors. However, coherent errors are always reversible in principle. Incoherent errors, in contrast, require quantum error correction to become reversible~\cite{Knill:1997kx}. Next, we present an overview of common incoherent and coherent errors in single- and two-qubit gates.

\subsection{Single-qubit gate errors}
In general the map $\mathcal{E}$ is very unconstrained. For a single qubit, i.e., for $D=2$, this allows $D^4-D^2=12$ real parameters in a CPTP map, and even more if we allow for leakage outside the qubit space (see below). 
On the other hand, the physical causes and impacts of error maps are more constrained. For a single qubit, the state $\rho$ can be represented by the Bloch vector $\mathbf{v}\equiv \left( \braket{\sigma_x},\braket{\sigma_y},\braket{\sigma_z} \right)$,
with the length of the vector giving the state purity. A general single-qubit unitary can be parametrized as $R(\vartheta,\theta,\phi)$, a rotation by an angle $\vartheta$ about an axis with polar and azimuthal angles $\theta$, $\phi$. Coherent errors in the space of qubit states can then be categorized as:
\begin{itemize}
    \item \emph{Rotation} errors, where the value of $\vartheta$ is changed from the target. These may be the result of drifts in the amplitudes of control drives or the coupling of the qubit to these drives.
    \item \emph{Phase} errors, where the axis of rotation is altered---that is, $\theta$ and/or $\phi$ are changed from the target. These may result from drifts in the qubit frequency leading to unwanted $z$-axis rotations, or drifts in the control axis (often caused by a drift in the phase of an oscillatory control field).
\end{itemize}
Similarly we can characterize incoherent errors as due to:
\begin{itemize}
    \item \emph{Relaxation} from the qubit state to the equilibrium state. These bit-flip errors are typically modeled as Markovian processes leading to an exponential decay with characteristic time $T_1$. Importantly, relaxation at finite temperature leads to a nonzero state purity as the system relaxes towards its ground state.
    \item \emph{Dephasing} due to noise in the qubit frequency relative to a reference (where the noise may originate from the qubit or the reference). These phase-flip errors may have both Markovian and non-Markovian components, and, in many condensed matter or solid-state systems, are typically modeled as a Markovian error plus a classical noise with a $1/f$ power spectrum~\cite{TripathiTransmon2024}. 
\end{itemize}

It is also important to consider \emph{leakage} errors, which result from the state evolving outside the qubit space. This error is coherent as long as it preserves state purity (thus allowing for its reversal~\cite{ByrdLidarWuZanardi:05}), although it will always appear incoherent if we restrict ourselves to only considering the qubit space. This is the case in, e.g., the transmon superconducting qubit~\cite{Koch2007}, where the qubit is encoded in the lowest two states of a weakly anharmonic oscillator. Driving this transition can also drive higher transitions, leading to coherent rotations in these state manifolds. Importantly, repeated application of a gate with a coherent leakage error can cause a rapid buildup of error that can be difficult to address. 

Modeling leakage in the qubit space requires a non-trace-preserving map, as population leaves the space and is transferred to the leakage state. Such maps can be modeled by modifying the CPTP map formalism (\cref{sec:QDM}) and replacing the equality in the $\sum_{i}A^{\dagger}A_{i}= I$ constraint by $\le I$. The corresponding process matrix $\bm{\chi}$ then has at most $D^4$
independent real elements ($16$ for a single qubit). 

A final error type, \emph{erasure}, is a special type of leakage error where qubit information is lost with some characteristic signature that an error has occurred. This is the case in, e.g., dual-rail qubit encodings where single-excitation loss takes both qubit states to a single leakage state and thus erases all information~\cite{Wu:2013vz,Levine:2024aa}, or trapped atom qubits where an atom may escape the trap~\cite{Grunzweig:2010aa}. Because erasure scrambles qubit information it must be considered as an incoherent error.
Some error correction architectures can correct erasure errors with far fewer resources than bit-flip or phase-flip errors. It can therefore be preferable to use encodings which transform other errors into erasures~\cite{Wu:2022aa}, as is done with dual-rail qubits. On the other hand, in many common qubits such as the transmon, erasure errors are not present and leakage must be considered as a coherent effect~\cite{Varbanov:2020aa,McEwen:2021aa}.

Incoherent errors are sometimes lumped together and modeled as a single depolarizing map 
\beq
\mathcal{E}(\rho)=(1-p)\rho+\frac{1}{3}p(X\rho X+Y\rho Y+Z\rho Z),
\label{eq:dep}
\eeq 
where $p$ is the probability of a qubit error. However, this both obscures their physical origin and does not account for the effects of biased noise\cite{gourlayConcatenatedCodingPresence2000,evansErrorCorrectionOptimisation2007,stephensAsymmetricQuantumError2008,Aliferis:2008aa,stephensHighthresholdTopologicalQuantum2013,brooksFaulttolerantQuantumComputation2013,XZZX2021} and finite temperature (see \cref{sec:T1asymm}).

If we ignore detection of erasure errors, and limit ourselves to Markovian decoherence, we see 7 parameters for qubit gate error: errors in $\vartheta, \theta, \phi$, relaxation rate, equilibrium population, dephasing rate, and leakage rotation. Some of these parameters (e.g., equilibrium population) may be independent of the gate itself, and so need not be characterized for every gate. Proper treatment of non-Markovian errors can be much more complicated~\cite{Kossakowski:10,Rivas:2014aa} and depends on the circumstance, but can often be captured with relatively simple models\cite{ShabaniLidar:05,zhangPredictingNonMarkovianSuperconductingQubit2022,TripathiTransmon2024}.

\subsection{Two-qubit gate errors}
The set of two-qubit gates is much larger and more diverse than the simple rotation gate for a single qubit, though it is amenable to an elegant geometrical classification~\cite{Zhang:03}. It is therefore more practical to consider the error maps for a particular two-qubit operation, and even for a particular physical implementation of that operation, rather than for general two-qubit operations. For instance, consider the parametrically-activated iSWAP gate in transmon qubits \cite{Sete2021}: an oscillatory flux drive causes oscillations of a tunable transmon's frequency about some maximum, bringing the average frequency into resonance with another transmon so the two can exchange energy (possibly via an intermediate tunable transmon coupler) \cite{seteErrorBudgetParametric2024}. Error maps include Markovian relaxation and dephasing of each qubit; non-Markovian dephasing of each qubit, typically modeled as a $1/f$ qubit frequency noise spectrum (note that the oscillatory nature of the drive acts as an ``auto-echo'' that eliminates much of this dephasing); two-qubit couplings to a Markovian or non-Markovian bath, e.g., a noisy tunable coupler; incorrect drive amplitude leading to transmon-transmon detuning and so causing incomplete swapping; two-qubit phase errors from unwanted $ZZ$ interaction due to higher transmon levels; single-qubit phase errors due to improper calibration of level repulsion during transients at the beginning and end of the gate; and leakage into higher transmon levels or coupler states. Every other implementation of iSWAP can have its own error maps, to say nothing of other two-qubit operations.

The complexity and diversity of two-qubit gate error maps mean it is best to approach them holistically on a case-by-case basis, e.g, by developing an ``error budget'' for the different sources of error~\cite{Tripathi2019, seteErrorBudgetParametric2024}. Another common approach is to characterize the gate completely once, using quantum process tomography or some other complete method. This identifies relevant error maps, which may then be measured more efficiently without needing to fully characterize the quantum process over and over. Typically the relevant coherent errors are characterized via some form of matrix element amplification, where repeated applications of the gate on a particular initial state build up a particular coherent error. There is a danger that this amplification will also amplify single-qubit coherent errors that may be difficult to distinguish. Adding dynamical decoupling sequences in between the two-qubit gate applications can help suppress these single-qubit errors that may confound the analysis\cite{Gross2024}. 

Incoherent errors in two-qubit gates are typically characterized by performing single-qubit coherence measurements, then using some error model (usually  Markovian) to predict the impact of single-qubit decoherence. Unfortunately, two-qubit incoherent errors due to joint bath couplings are often ignored unless full process tomography is being performed. This simplification can be shown to give good predictions in many contexts\cite{gaikwadEntanglementAssistedProbe2024}, but in some circumstances two-qubit incoherent errors may be quite significant. For instance, the two qubits are often coupled via a third system (such as a tunable coupler) that can act as a joint bath. Fortunately, it is straightforward to extend standard single-qubit coherence measurements---for instance, joint relaxation can be characterized by comparing the relaxation rates of the $\ket{10},\ket{01}$, and $\ket{11}$ states \cite{DubiRelaxation2009}.

Properly benchmarking two-qubit gates is a significant challenge, and the best approach depends on the exact goal: calibrating a gate, predicting algorithmic performance, characterizing the overall quality of a quantum processor, etc. We describe and compare the various single- and two-qubit gate benchmarking protocols below.

\section{Gate and Circuit Benchmarking Protocols}
\label{sec:protocols}

A gate benchmarking protocol produces a set of numbers for an input gate or gate set, which can be used to compare the performance in different instances of gate implementation. Such a protocol can be judged on the basis of a few main criteria:
\begin{itemize}
\item Comprehensiveness, i.e., how much information about the error does the protocol provide. One simple way to quantify the amount of information provided is the number of parameters that end up describing the error. 
\item Specificity, i.e., what assumptions are made about the noise model in the protocol. Some common assumptions are that errors are independent of each other and weakly depend on the gate being applied\cite{Greenbaum_2015,Brown_Eastin_2018}, or that they are of a particular type, such as Pauli errors~\cite{seif2024entanglementenhancedlearningquantumprocesses}.
\item Simplicity, i.e., how easily the scheme can be adapted to existing experimental setups. This includes how complicated the pulse control and measurement schemes are for the experiment as well as how sensitively the experiment must be calibrated.
\item Resource costs, i.e., how expensive it is to run the protocol. Note that scalability, which is a critical consideration in quantum algorithms and processes, can become less relevant for a gate benchmarking protocol if we restrict the characterization effort to single 1 or 2 qubit gates in isolation. The number of such gates is at most $n(n-1)/2$ for $n$ fully connected qubits. Instead, the absolute resource cost of the protocol becomes important and needs to be optimized. 
\end{itemize}
Apart from these criteria, a gate benchmarking protocol should also be robust against all errors, apart from the ones it is trying to quantify. Critically, for most protocols, this robustness should include state preparation and measurement (SPAM) errors~\cite{Khan:2024aa}, which are among the most pervasive errors in experiments\cite{PhysRevA.87.062119}. Exceptions to this rule include protocols that quantify the SPAM errors themselves.

\subsection{Randomized Benchmarking}
\label{sec:rb} 
Randomized Benchmarking (RB) aims to give the most concise characterization of gate error by reporting a single number to quantify the average gate fidelities, while suppressing the effects of SPAM errors \cite{Emerson_2005, Knill2008, Magesan2011, Magesan2012}. Over the past two decades, RB has become a standard tool used across different platforms to quantify the performance of single- and two-qubit gates. Many variants of RB have been introduced; here, we focus on the two most prominent variants, known as standard RB and interleaved RB. The main benefits of RB methods include reasonable assumptions, simplicity of implementation compared to process tomography, and relatively low resource cost.
However, RB reduces the characterization to a single number. Indeed, this is the main reason for the method's relative simplicity.

\subsubsection{Standard RB}
\label{sec:SRB}

Standard RB characterizes a specific selection of gates, which are elements of the Clifford group (see below), by an average gate infidelity metric. The key idea behind standard RB is to transform errors into an effective depolarizing error [recall \cref{{eq:dep}}] via a process called ``twirling", which simplifies the analysis. 
The process is described in detail below.

The Pauli group $\mathcal{P}_n$ is the group of all possible $n$-fold tensor products of the three Pauli matrices and the identity matrix. The Clifford group consists of the set of gates that preserve the Pauli group under conjugation. I.e., $C$ is in the Clifford group if $CPC^\dag\in\mathcal{P}_n$ for all $P\in\mathcal{P}_n$.

The standard RB protocol involves the following steps:
\begin{itemize}

\item \textit{Preparation}: Initialize the $n$-qubit quantum system in a reference state $|0\rangle ^{\otimes n}$, typically the ground state. 

\item \textit{Random Gate Sequence}: Apply a sequence of $m$ randomly selected Clifford gates $\{C_i\}_{i=1}^m$:
\begin{equation}
U_m = C_m C_{m-1} \cdots C_1.
\end{equation}

\item \textit{Inversion Gate}: Apply the inverse of the accumulated operation to ideally return the system to the initial state:
\begin{equation}
C_{\text{inv}} = C_{m+1} =  U_m^\dagger = C_1^\dagger \cdots C_m^\dagger.
\end{equation}
The inverse can be calculated efficiently classically~\cite{PhysRevA.70.052328}.
\item \textit{Measurement}: Measure the system in the computational basis (the eigenbasis of the Pauli $\sigma_z$ matrix) and record whether it has returned to the initial state.
\item \textit{Averaging}: Average the measurement outcomes over $K$ different random sequences for a given $m$ to obtain the survival probability $P(m)$. Repeat the process for various sequence lengths $m$ to get a sequence of survival probabilities $\{P(m)\}$.

\item \textit{Fitting to exponential}: The final step is to fit the measured survival probability to an exponential function $Ap^m + B$ in order to extract $p$, which can be used to quantify the \emph{average gate fidelity} of the Clifford gate sequences. Here, $A$ and $B$ capture the SPAM errors. 
\end{itemize} 

\subsubsection{RB Theory}
A randomly selected Clifford gate $C$ can be represented by the ideal superoperator $\mathcal{G}(\cdot) = C (\cdot)C^\dagger$, and the associated error process is modeled as $\tilde{\mathcal{G}} = \mathcal{E} \circ \mathcal{G}$, where the error $\mathcal{E}$ is assumed to be deterministic.  The total process map $\tilde{\mathcal{I}}_m$ for the sequence of $m+1$ Clifford gates (including the last one which inverts the previous $m$ gates) is given by:
\begin{equation}
    \tilde{\mathcal{I}}_m = \tilde{\mathcal{G}}_{m+1} \circ\tilde{\mathcal{G}}_m \circ \cdots \circ \tilde{\mathcal{G}}_1 .
\end{equation}
Ideally, $\tilde{\mathcal{I}}_m$ should have been the identity operation. Assuming gate-independent noise i.e., $\mathcal{E}_i \equiv \mathcal{E}$ $\forall i$, we obtain:
\begin{equation}
    \tilde{\mathcal{I}}_m = \left( \mathcal{E} \circ \mathcal{G}_{m+1} \right) \circ \left( \mathcal{E} \circ \mathcal{G}_{m} \right) \circ \cdots \circ \left( \mathcal{E} \circ \mathcal{G}_1 \right)
\end{equation}
which can be written as
\begin{equation} 
\tilde{\mathcal{I}}_m = \mathcal{E}\circ \bigcirc_{i=1}^{m} \left( \mathcal{D}_{i}^{-1} \circ \mathcal{E} \circ \mathcal{D}_{i} \right) \end{equation}
where $\mathcal{D}_i = \prod_{k=1}^{i} \mathcal{G}_{k}$ and we have used $\Pi_{k=1}^{m+1}\mathcal{G}_k = I$. This is the process generated by one of the realizations; repeating the same depth-$m$ sequence $K$ times results in an average process defined as
\begin{equation}
\begin{split} 
\overline{\mathcal{I}}_m &\equiv \frac{1}{K}\sum_{j}\tilde{\mathcal{I}}_{m}^{(j)}\\
&=  \mathcal{E}\circ \frac{1}{K}\sum_{j} \bigcirc_{i=1}^{m} \left( \mathcal{D}_{i}(j)^{-1} \circ \mathcal{E} \circ \mathcal{D}_{i}(j)\right).
\end{split}
\end{equation}
Here the error map $\mathcal{E}$ is ``twirled'' over the Clifford group, and it can be shown~\cite{Nielsen:2002aa} that this reduces it to an $m$-fold depolarizing map $\mathcal{E}_{\rm dep}^{\circ m}$
which replaces a state $\rho$ by the maximally mixed state $I/D$ with probability $1-p^m$, and leaves the state unchanged with probability $p^m$. We note that this process of applying random unitaries that leave only a Markovian contribution is closely related to the earlier idea of randomized dynamical decoupling~\cite{Viola:2005:060502,Santos:05,Santos:2006:150501}.

The average survival probability of the initial state $\rho_0 = (\ketb{0}{0})^{\otimes n}$ is then given by:
\begin{equation}
\label{eq:RB_decay}
\begin{split}
    P_{\rm avg}(m)&= {\rm Tr}[\rho_0 ~\overline{\mathcal{I}}_m(\rho_0)]\\
    &= {\rm Tr}[\rho_0 \mathcal{E} \circ \mathcal{E}_{\rm dep}^{\circ m} (\rho_0)]\\
    &= {\rm Tr}\left[\rho_0 \mathcal{E}  \left(p^m \rho_0 + (1-p^m) \frac{I}{D}\right)\right]\\
    &= A p^m+ B  \end{split}
\end{equation}
where 
\begin{equation}
    A = {\rm Tr}\left[\rho_0 \mathcal{E}\left(\rho_0 - \frac{I}{D}\right)\right] ,
\end{equation}
and
\begin{equation}
B = {\rm Tr}\left[\rho_0 \mathcal{E}\left(\frac{I}{D}\right)\right] .
\end{equation}

The key assumptions include gate-independent and time-independent errors in the above derivation. 
$P_{\rm avg}(m)$ from \cref{eq:RB_decay} is used to fit the experimental decay curves to extract $p$. Since twirling generates a depolarizing noise map of the form $\mathcal{E}_{\rm dep}(\rho) = p \rho + (1-p)\frac{I}{D}$, its average gate fidelity with respect to an identity  can be calculated to be $\overline{F}_{\mathcal{E}_{\rm dep}, I} = p+(1-p)/D$, where the depolarization parameter $p$ is identified with the experimentally measured exponential decay parameter. Therefore, we can relate the average Clifford gate infidelity $r_C = 1 - \overline{F}_{\mathcal{E}_{\rm dep}, I}$ to the experimental decay parameter $p$ obtained from the RB experiments as:
\begin{equation}
r_C= \frac{D-1}{D}(1-p)  .
\label{eq:rb_infid}  
\end{equation}

Since $r_C$ depends only on the depolarization parameter $p$ of the effective process matrix generated by the gate, RB becomes insensitive to SPAM errors, which are captured by $A$ and $B$.

If we consider the case when the errors are gate-dependent as well as time-dependent, a perturbation theory calculation can be performed to extract the higher order corrections. In particular, the first order correction yields \cite{Magesan2011,Magesan2012}:

\begin{equation}
    P^{(1)}_{\rm avg}(m) = A_1 p^m + B_1 + C_1 (m-1)(q-p^2)p^{m-2}\label{eq:second-order}
\end{equation}
where $A_1$, $B_1$ and $q$ are functions of $m$. The term $(q-p^2)$ is a measure of the degree of gate-dependence in the errors. It can be shown  
that the second order term can be neglected when  $\gamma \ll 2/m$, where $\gamma$ is a measure of the noise variation.~\cite{Magesan2011} 
The zeroth order approximation [see \cref{{eq:RB_decay}}] provides a good estimate for small errors and the first order correction may be needed in cases where the errors are large. Note that even if the gate dependence term $q-p^2$ is large, it would still be possible to fit the experimental data with \cref{eq:second-order} as long as $m$ is large enough. However, this can conflict with the $\gamma \ll 2/m$ condition. In addition, it is also necessary for $m$ to be large enough to have enough data points for a
reasonable estimate of $p$ in the fitting model. These considerations highlight some of the restrictions of using RB for gate and time-dependent noise.

\subsubsection{Interleaved RB}
\label{sec:IRB}

Whereas RB provides an averaged fidelity for a random set of Clifford gates, \emph{Interleaved RB} (IRB)~\cite{Magesan:2012aa} allows the estimation of the fidelity of a specific Clifford gate $\mathcal{C}$. The key idea is to analyze the effect of inserting the gate of interest $G$ (represented below by a superoperator $\mathcal{G}$) into an RB sequence. The IRB protocol consists of two different experiments: the first is the standard RB as discussed in \cref{sec:SRB}, while the second is an RB sequence with $\mathcal{G}$ interleaved between every pair of consecutive random Clifford gate. More concretely, the sequence of gates in the second experiment after interleaving is:

\begin{equation}
    {\mathcal{U}}_{m,\rm inter} = {\mathcal{G}}_m \circ\mathcal{G} \circ{\mathcal{G}}_{m-1} \circ \cdots \circ\mathcal{G}\circ {\mathcal{G}}_2 \circ\mathcal{G}\circ {\mathcal{G}}_1 .
\end{equation}
The above sequence is then followed by $\mathcal{G}_{m+1} = {\mathcal{U}}_{m,\rm inter}^{-1}$. Similar to standard RB, this also generates an exponential decay characterized by a parameter $p_{\rm inter}$. It can then be shown~\cite{Magesan:2012aa} that using the two decay parameters $p$ and $p_{\rm inter}$ from the two experiments, we can estimate the gate infidelity contribution of the particular gate $\mathcal{G}$ using:
\begin{equation}
\label{eq:r_interleaved}
r_{\mathcal{G}} = \frac{D-1}{D}\left(1-\frac{p_{\rm inter}}{p} \right) .
\end{equation}

We note that IRB is not always a reliable method for estimating a gate's infidelity because unitary errors in the interleaved gate $\mathcal{G}$ can coherently add or cancel with errors in the random Clifford gates, sometimes even resulting in a negative error rate~\cite{hashim2024}. Another way to see the problem with IRB is via \cref{eq:r_interleaved} with the interleaved decay slower than the decay of standard RB ($p_{\rm inter}>p$), which results in a negative infidelity $r_{\mathcal{G}}$. This, in turn, implies an unphysical average gate fidelity larger than $1$. Such issues lead to potentially large discrepancies between the estimated and true fidelity.

In fact, it can be shown \cite{Magesan:2012aa} that the original error rate for the Clifford gate of interest $\mathcal{G}$ must lie in the range $[r_{\mathcal{G}} - E, r_{\mathcal{G}} + E]$ where

\begin{equation}
E=\min \left\{\begin{array}{l}
\frac{(D-1)\left[\left|p-p_{\rm{inter}} / p\right|+(1-p)\right]}{D} \\
\frac{2\left(D^2-1\right)(1-p)}{p D^2}+\frac{4 \sqrt{1-p} \sqrt{D^2-1}}{p}
\end{array}\right. ,
\end{equation}
which arises from imperfect random gates. When the noise channel is fully depolarizing ($p=1$) then $E=0$, but in general $E>0$. This means that $r_{\mathcal{G}} + E$ can exceed $1$, which is unphysical. When this happens, IRB results must be treated with caution. Possible remedies include increasing the number of random sequences and the number of circuit repetitions, and checking for systematics in how errors are introduced (e.g., calibration or readout errors might be overshadowing gate error in unexpected ways), 

Despite its shortcomings, randomized benchmarking is a fundamental tool in quantum computing for quantifying average gate fidelities with robustness to SPAM errors. By leveraging the mathematical properties of the Clifford group and statistical averaging, RB provides a scalable method for characterizing quantum processors. Ongoing research continues to refine RB protocols and extend their applicability to more general settings. IRB is just one possible extension of RB; there is a plethora of other work focusing on other generalizations of randomized benchmarking. These include simultaneous RB \cite{Gambetta2012} (which detects the effects of simultaneous gate operations), leakage RB \cite{Wallman_2016,Wood2018}, non-Markovian RB \cite{Fogarty2015, Wallman2018randomized}, mirror RB \cite{Proctor2022mirror}, character RB \cite{Helsen2019, Helsen2022} (which goes beyond the Clifford group), and more.

\subsection{Quantum Process Tomography}

If randomized benchmarking sacrifices some comprehensiveness and specificity in service of simplicity and low resource cost, quantum process tomography (QPT) does the opposite. This method seeks to fully characterize a quantum process $\mathcal{E}$, instead of providing a single number as in RB. Since $D^4-D^2$ real numbers are required to fully specify a CPTP map $\mathcal{E}$, QPT is, of course, much more resource intensive than RB.

\subsubsection{Standard Quantum Process Tomography}
\label{SQPT-sec}

\begin{figure}[tp]
\centering
\includegraphics[width=0.8\columnwidth]{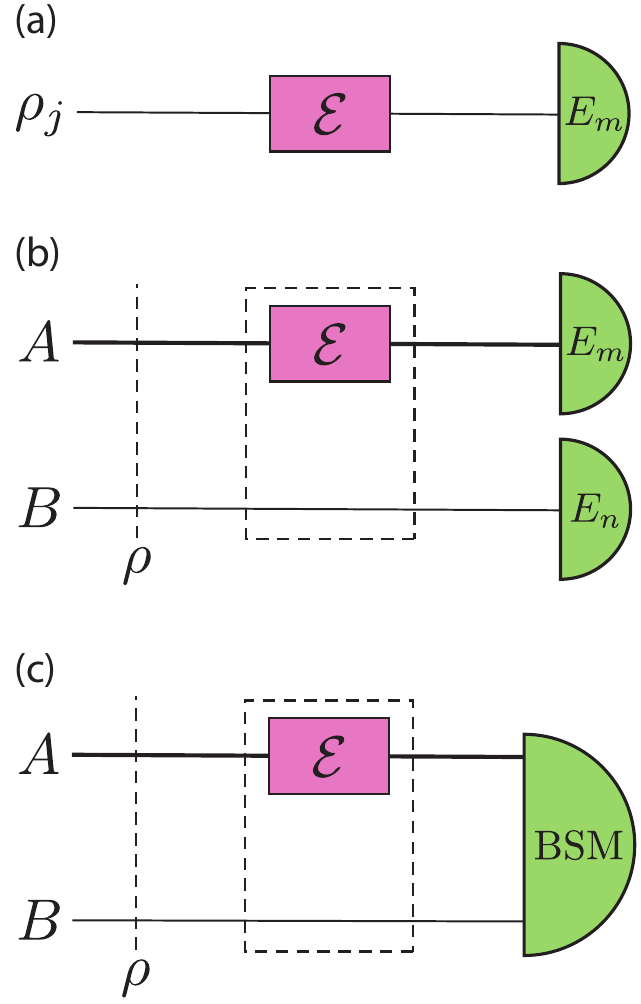}
\caption{Three different QPT methods. (a) Schematic of SQPT. An ensemble of states $\{\protect\rho_j\}$ is prepared and each state is subjected to the map $\mathcal{E}$ followed by measurements $\{E_m\}$. (b) Schematic of separable AAPT. A ``faithful" input state $\protect\rho$ is subjected to the map $\mathcal{E}\otimes I$. The operators $\{E_{j}\}$ are measured on the joint system, which results in the required joint probability distributions or expectation values. (c) Schematic of DCQD. The system and the ancilla are prepared in one of the input states as in \cref{dcqd-tab} and is subjected to the map $\mathcal{E}\otimes I$. The joint system is measured in the Bell-state basis (Bell State Measurement).}
\label{fig:QPT}
\end{figure}

In ``standard'' quantum process tomography (SQPT), illustrated in \cref{fig:QPT}(a), we prepare $D^2$ linearly independent inputs $\{\rho _{k}\}_{k=0}^{D^2-1}$, where each
$\rho_k \in \mathcal{B}(\mathcal{H})$,
and then measure the output states $\mathcal{E}(\rho _{k})$ using quantum state tomography~\cite{chuang-sqpt,poyatos-sqpt}. Since the map $\mathcal{E}$ [\cref{eq:E}] is linear, it can be in principle reconstructed from the measured data by proper inversion. Let
$\{\rho _{k}\}_{k=0}^{D^2-1}$ be a linearly independent basis set of
operators for $\mathcal{B}(\mathcal{H})$. A convenient choice
is $\rho _{k}=\ketb{m}{n}$, where $\{\ket{m} \}_{m=0}^{D-1}$ is
an orthonormal basis for $\mathcal{H}$ and $k=(m,n)$. The coherence $\ketb{m}{n}$
can be constructed from four populations:
\begin{align*} \ketb{m}{n} &= \ketb{+_{mn}}{+_{mn}} + \ketb{-_{mn}}{-_{mn}} \\
&\qquad - \frac{1+i}{2}\left(\ketb{m}{m}+\ketb{n}{n}\right),
\end{align*}
where $\ket{\pm_{mn}} = (\ket{m}\pm\ket{n})/\sqrt{2}$. The linearity of $\mathcal{E}$
then implies that a measurement of $\mathcal{E}(\ketb{+_{mn}}{+_{mn}})$, $%
\mathcal{E}(\ketb{-_{mn}}{-_{mn}})$, $\mathcal{E}(\ketb{m}{m})$, and
$\mathcal{E}(\ketb{n}{n})$ suffices for the determination of $%
\mathcal{E}(\ketb{m}{n})$. In addition, the channel output $\mathcal{E}(\rho_k)$ itself can be expressed in terms of a linear combination of basis states, as
$\mathcal{E}(\rho_k)=\sum_{l}\lambda _{kl}\rho _{l}$, where the parameters 
\begin{equation}
\lambda _{kl}=\Tr (\rho_{k}\mathcal{E}(\rho _{l})) 
\label{eq:lambda_kl}
\end{equation}
contain the measurement results. Similarly, each summand in RHS of \cref{eq:chi-from-E} can be expanded in the basis states as $E_{m}\rho_{k}E_{n}^{\dagger}=\sum_{l}B_{mn,lk}\rho _{l}$. The resulting equation then reads
\begin{equation}
\begin{split}
    \mathcal{E}(\rho_k) = \sum_{mnl}B_{mn,lk}\rho _{l} &= \sum_{l}\lambda _{kl}\rho_{l} \\
    \iff \sum_{mn}B_{mn,lk}\chi _{mn}&=\lambda _{kl}
\end{split}
\end{equation}
where the second line follows from the linear independence of $\{\rho_k\}$. This relation can in turn be written in
the following matrix form:
\begin{equation}
\bm{B}\bm{\chi}=\bm{\lambda},  \label{sqpt}
\end{equation}%
where the $D^{4}\times D^{4}$-dimensional matrix $\bm{B}$ is determined by
the choice of operator bases $\{\rho _{k}\}$ and $\{E_{m}\}$, and the $D^{4}$-dimensional vector $\bm{\lambda}$ is determined from state tomography experiments. The process matrix $\bm{\chi}$ can thus be determined by inversion of \cref{sqpt}.

In general, SQPT involves the preparation of $D^2$ linearly independent inputs $\{\rho _{l}\}$, each of which is subjected to the quantum process $\mathcal{E}$, followed by quantum state tomography on the corresponding outputs. For each $\rho_{l}$, we must measure the expectation values of the $D^2$ fixed-basis operators $\{E_{k}\}$ in the state $\mathcal{E}(\rho _{l})$. Thus, the total number of measurements required is $D^4$. Since measurement of an expectation value cannot be done on a single copy of a system, whenever we use the term \textquotedblleft
measurement\textquotedblright\ we implicitly mean measurement on an \emph{ensemble} of identically prepared quantum systems corresponding to a given experimental setting.


\subsubsection{Ancilla-Assisted Process Tomography}
\label{AAPT-sec}

There is a duality between quantum
\textit{state} tomography schemes and QPT. This duality is based
upon the Choi-Jamiolkowski isomorphism
\cite{Jamiolkowski:72,Choi:75} we discussed in \cref{sec:QDM}, which establishes a correspondence
between CP maps and states.
This one-to-one correspondence enables all theorems about quantum
maps to be directly derived from those of quantum states
\cite{Arrighi:2004aa}. In particular, 
it allows one to represent a quantum process as a quantum state in a larger Hilbert space. For example, choosing
$\{E_m\}=\{\ketb{i}{j}\}$ results in
$\bm{\chi}=D\rho_{\mathcal{E}}$, where $\rho_{\mathcal{E}}$ is the Choi state of the CPTP map $\mc{E}$.
Therefore, QPT has a natural reduction to quantum state tomography, and state identification methods can be applied to the characterization of quantum processes as well. The ``ancilla-assisted process tomography''
(AAPT)~\cite{DArianoAAQPT,Leung:03,altepeter-aapt} scheme is based on this observation and is schematically illustrated in \cref{fig:QPT}(b).

The AAPT scheme involves attaching an ancilla $B$ to the principal system $A$ and preparing the combined system in a particular joint state so that complete information
about $\mathcal{E}$ can be imprinted on the final state. Then, by performing quantum
state tomography in the extended Hilbert space $\mathcal{H}_{AB}=\mathcal{H}_{A}{\otimes }\mathcal{H}_{B}$, one can extract complete information about the
unknown map $\mathcal{E}$ acting on $A$. In principle, the input
state of the system and ancilla can be prepared in either an
entangled mixed state (entanglement-assisted) or a separable mixed
state. The only requirement for the input state in AAPT is that it must be ``faithful'' 
to the map $\mathcal{E}$, i.e., performing quantum state tomography on the output identifies $\mathcal{E}$ completely and
unambiguously~\cite{dariano-faithful}. The faithfulness condition implies that a state $\rho $ can
be used as input for AAPT iff it has maximal Schmidt number $D^2$, which is essentially an invertibility condition. By performing state tomography on the output state $(\mathcal{E}\otimes I)(\rho )$, we obtain a set of linear relations among
possible measurement outcomes and the elements of $\mathcal{E}$ and $\rho$; the choice of $\rho $ must be such that an inversion is possible.

Maximal entanglement is not a necessary property of the input state $\rho $ in AAPT.
However, there is experimental evidence that maximally entangled pure states offer
optimal performance in the sense of reducing experimental error~\cite{altepeter-aapt}. Indeed, experimental error amplification is related to inversion of the matrix involving $\mathcal{E}$ and $\mathcal{\rho}$, which, in turn, depends on the magnitude of the inverse of the eigenvalues of $\rho$. Thus, a smaller
eigenvalue leads to an amplified experimental error. As a consequence, the optimal (in the sense of minimal experimental errors) faithful input states are maximally entangled pure states.

One can realize the required quantum state tomography via separable measurements
(separable AAPT), i.e., joint measurement of tensor product operators, or via
collective measurements on both the system and the ancilla (non-separable AAPT).
Both types of measurements are performed on the same Hilbert space $\mathcal{%
H}_{AB}$. Furthermore, by adding another ancilla it is possible to perform a generalized measurement on the joint system $AB$. It turns out that the non-separable
measurement schemes (whether in the same Hilbert space or in a larger one)
hardly offer any practical advantage because they
require many-body interactions which are experimentally unrealizable~\cite{Mohseni:2008ly}. We thus focus on joint separable measurements.

Let us assume that the initial state of the system and the ancilla is $\rho
_{AB}=\sum_{ij}\rho _{ij}E_{i}^{A}\otimes E_{j}^{B}$, where $\{E_{m}^{A}\}$ (%
$\{E_{n}^{B}\}$) is the operator basis for the linear operators acting on $%
\mathcal{H}_{A}$ ($\mathcal{H}_{B}$). The output state,
after the action of the unknown map $\mathcal{E}$ on the principal system, is:
\begin{equation}
\begin{split}
\label{eq:AAPT}
\rho _{AB}^{\prime } =(\mathcal{E}_{A}\otimes I_{B})(\rho _{AB})
=\sum_{kj}\tilde{\alpha}_{kj}E_{k}^{A}\otimes E_{j}^{B} , 
\end{split}
\end{equation}
where $\tilde{\alpha}_{kj}\equiv\sum_{mni}\chi _{mn}\rho
_{ij}\alpha _{k}^{m,i,n}$, with $\alpha _{k}^{m,i,n}$ defined via $E_{m}^{A}E_{i}^{A}E_{n}^{A\dagger }=\sum_{k}\alpha _{k}^{m,i,n}E_{k}^{A}$. Note that $\alpha _{k}^{m,i,n}$ depends only on the choice of operator basis. 
The result of measuring the observables $E_{k}^{A}\otimes E_{j}^{B}$ is:
\begin{equation}
\widetilde{\alpha }_{kj}=\Tr (\rho _{AB}^{\prime }(E_{k}^{A
}\otimes E_{j}^{B })^{\dagger}).  \label{eq:JSM}
\end{equation}
Now, let
\begin{equation}
\widetilde{\chi }_{ki}=\sum_{mn}\alpha _{k}^{m,i,n}\chi _{mn},  \label{chi}
\end{equation}%
where the process matrix $\bm{\chi}$ is the primary objective of QPT [recall \cref{eq:chi-from-E}] and the $\alpha $ parameters are known from the choice of
operator basis. Thus, the procedure to obtain $\bm{\chi}$ simplifies to two matrix inversions: after measuring $\{\widetilde{\alpha }_{kj}\}$, we obtain $\widetilde{\bm{\chi}}$ by inverting the following matrix equation:
\begin{equation}
\widetilde{\bm{\alpha}}=\widetilde{\bm{\chi}}~\bm{\varrho}.
\label{matrix}
\end{equation}
Using $\widetilde{\bm{\chi}}$, $\bm{\chi}$ is then found by inverting \cref{chi}.

\subsubsection{Direct Characterization of Quantum Dynamics}

\label{DCQD-sec}

An alternative to SQPT and AAPT is \textquotedblleft direct characterization of
quantum dynamics\textquotedblright\ (DCQD)
\cite{MohseniLidar:06,MohseniLidar:07,Mohseni:2008ly,MohseniHI,Graham:2013aa,Nigg:2013aa}. In DCQD, similarly to AAPT,
an ancilla system is utilized, but in stark
contrast, it does not require inversion of a $D^2\times D^2$
matrix; hence, ``direct". The main idea in DCQD is to use certain entangled states
as inputs and to perform a simple error-detecting measurement on the
joint system-ancilla Hilbert space. The choice of input
states and measurements is such that the measurement results directly encode elements of the quantum map, which
removes the need for state tomography. In the case of a single qubit, the
measurement scheme turns out to be equivalent to a Bell state
measurement (BSM). In DCQD the choice of input states is dictated by
whether diagonal or off-diagonal elements (quantum dynamical population and coherence, respectively) of the process matrix are to be determined. Population
characterization requires maximally entangled input states, while
coherence characterization requires nonmaximally entangled input
states. We review the scheme for the case of
a single qubit below. 

Let us first demonstrate how to determine all diagonal elements of the process matrix, $\{\chi _{mm}\}$, in a single
(ensemble) measurement. We choose $\{I,X^{A},Y^{A},Z^{A}\}$ as our
operator basis acting on the principal qubit $A$. We first prepare the maximally
entangled Bell state $|\Phi ^{+}\rangle =(|00\rangle +|11\rangle )_{AB}/\sqrt{2}$, and then (as in AAPT) subject only qubit $A$ to the map $\mathcal{E}$.

A stabilizer code is a subspace $\mathcal{H}_{C}$ of the Hilbert space of $n$
qubits that is an eigenspace of a given Abelian subgroup $\mathcal{S}$ that excludes $-I$ (the
stabilizer group) of the $n$-qubit Pauli group $\mathcal{P}_n$, with eigenvalue $+1$ \cite{Gottesman:1996fk}. In other words, for every $|\Psi
_{C}\rangle \in \mathcal{H}_{C}$ and all $S_{i}\in \mathcal{S}\subset \mathcal{P}_n$, we have $S_{i}|\Psi _{C}\rangle =|\Psi _{C}\rangle $, where $[S_{i},S_{j}]=0$ $\forall i,j$. Consider
the action of an arbitrary (error) operator $E\in\mathcal{P}_n$ on a stabilizer code state
$|\Psi _{C}\rangle$. Such an error is detectable if it anticommutes with (at least one of) the stabilizer generators: $\{S_{i},E\}=0$. To see this note that
\begin{equation}
S_{i}(E|\Psi _{C}\rangle )=-E(S_{i}|\Psi _{C}\rangle )=-(E|\Psi _{C}\rangle
),
\end{equation}
i.e., $E|\Psi _{C}\rangle $ is a $-1$ eigenstate of $S_{i}$. Hence, measuring $S_{i}$ detects the occurrence of an error or no error (the $-1$ or $+1$ outcomes respectively). Measuring all the stabilizer generators then yields a list of errors (the \textquotedblleft syndrome\textquotedblright), which allows one to detect the errors that took place. More precisely, each stabilizer code has a distance parameter $d$. The code detects all errors whose weight (number of non-identity operators) is at most $d-1$. E.g., a distance $d=3$ code can detect all one- and two-body errors, i.e., errors of the form $XII$, $IYZ$, etc., but not weight-$3$ errors such as $XYZ$.

The Bell state $\ket{\Phi ^{+}} $ is a $+1$ eigenstate of the commuting
stabilizer generators $Z^{A}Z^{B}$ and $X^{A}X^{B}$, i.e., it is stabilized under their action. The non-identity error operators in the operator basis set $\{E_{m}^A\}_{m=0}^3= \{I,X^{A},Y^{A},Z^{A}\}$ anticommute with at least one of the stabilizer
generators, and therefore by measuring these generators we can detect the
error.
Note that, as mentioned above, measuring $Z^{A}Z^{B}$ and $X^{A}X^{B}$ is
equivalent to a BSM: it can be represented by the four projection operators
$P_{\Psi ^{\pm }}=\ketb{\Psi ^{\pm }}{\Psi ^{\pm }} $
and $P_{\Phi ^{\pm }}=\ketb{\Phi ^{\pm }}{\Phi ^{\pm }}$, where $%
\ket{\Phi ^{\pm }} =(\ket{00}\pm \ket{11} )/\sqrt{2}$ , and $\ket{\Psi ^{\pm }} =(\ket{01}\pm \ket{10} )/\sqrt{2}$ are the Bell states.
Expressing the map on qubit $A$ as $\mathcal{E}(\rho) = \sum_{m,n}\chi_{mn}E_m^A\rho E_n^{A\dagger}$, it is easily verified that the probabilities of obtaining the no-error outcome $I$ and the errors $X^{A}$, $Z^{A}$, and $Y^{A}$ can be written as:
\begin{equation}
p_{m}=\Tr [P_{m}(\mathcal{E}\otimes I)(\ketb{\Phi ^{+}}{\Phi ^{+}})]=\chi _{mm},  
\label{eq:pm}
\end{equation}
where $m=0,1,2,3$, and the projectors $P_{m}$, for $m=0,1,2,3$,
correspond to the states $\Phi ^{+}$, $\Psi ^{+}$, $\Psi ^{-}$, and $\Phi ^{-}$, respectively. For example:
\begin{equation}
    \begin{split}
        &\Tr [P_{0}(\mathcal{E}\otimes I)(\ketb{\Phi ^{+}}{\Phi ^{+}})] \\
        &\quad = \sum_{m,n=0}^3\chi_{mn}\Tr [P_{\Phi ^{+}}(E_m^A\otimes I) P_{\Phi ^{+}}(E_n^{A\dagger}\otimes I)] \\
        &\quad = \sum_{m,n=0}^3\chi_{mn} \delta_{m0} \delta_{n0} = \chi_{00} 
    \end{split}
\end{equation}
\cref{eq:pm} shows that the diagonal elements of the
process matrix are directly obtainable from an ensemble BSM; this is
the core observation that leads to the DCQD scheme. In particular,
one can determine the quantum dynamical populations, $\chi _{mm}$, by simultaneously measuring the
operators $Z^{A}Z^{B}$ and $X^{A}X^{B}$) on multiple copies of the
state $|\Phi ^{+}\rangle$.

\begin{table*}[tp]
\begin{tabular}{c|ccc|c}
\multicolumn{1}{c|}{input state}&\multicolumn{3}{c|}{Measurement}&\multicolumn{1}{c}{Output}\\
\hline
 & Stabilizer & Normalizer & Bell state measurement& $\chi_{mn}$\\
\hline
$\ket{\Phi^+}$& $Z^AZ^B,X^AX^B$ & N/A & $P_{\Psi^{\pm}},P_{\Phi^{\pm}}$ & $\chi_{00},\chi_{11},\chi_{22},\chi_{33}$\\
$\ket{\Phi^+_\alpha}$ & $Z^AZ^B$ & $X^AX^B$ & $P_{\Phi^{+}}\pm P_{\Phi^{-}}, P_{\Psi^{+}}\pm P_{\Psi^{-}}$ & $\chi_{03},\chi_{12}$\\
$\ket{\Phi^+_\alpha}_X$ & $X^A X^B$ & $Z^A Z^B$ & $P_{\Phi^{+}}\pm P_{\Psi^{+}},P_{\Phi^{-}}\pm P_{\Psi^{-}}$ & $\chi_{01},\chi_{23}$\\
$\ket{\Phi^+_\alpha}_Y$ & $Y^A Y^B$ & $Z^A Z^B$ & $P_{\Phi^{+}}\pm P_{\Psi^{-}},P_{\Phi^{-}}\pm P_{\Psi^{+}}$ & $\chi_{02},\chi_{13}$\\
\end{tabular}
\caption{Input states and measurements for
direct characterization of quantum dynamics (DCQD) for a
single qubit. Here
$|\Phi^{+}_{\alpha}\rangle=\alpha|00\rangle+\beta|11\rangle$
($|\alpha|\neq 0,1/\sqrt{2}$),
$\ket{\Phi^{+}_{\alpha}}_{X(Y)}=\alpha\ket{++}_{X(Y)}+\beta\ket{--}_{X(Y)}$
($|\alpha|\neq 0,1/\sqrt{2}$ and
$\text{Im}(\alpha\beta^*)\neq0$) and $\{|0\rangle,|1\rangle \}$,
$\{|\pm\rangle_X \}$, $\{|\pm\rangle_Y\}$ are eigenstates of the
Pauli operators $Z$, $X$, and $Y$. The fourth column shows the Bell state
measurement (BSM) equivalent of the stabilizer + normalizer measurements. The last column is the output: all elements of the process matrix $\bm{\chi}$.}
\label{dcqd-tab}
\end{table*}

To determine the quantum dynamical coherence elements, $\chi _{m\neq n}$, a different strategy
is needed. As the input state we take a non-maximally entangled state: $%
|\Phi _{\alpha }^{+}\rangle =\alpha |00\rangle +\beta |11\rangle $, with $%
|\alpha |,|\beta |\notin \{0,1/\sqrt{2}\}$ and
$\text{Im}(\alpha\beta^*)\neq 0$. The sole stabilizer of this
state is $Z^{A}Z^{B}$. It is simple to verify that by measuring $Z^{A}Z^{B}$
on the output state $\mathcal{E}(\rho )$, with $\rho =|\Phi _{\alpha
}^{+}\rangle \langle \Phi _{\alpha }^{+}|$, we obtain the following probabilities for $\pm1$ outcomes:
\begin{equation}
\begin{split}
\Tr [P_{+1}\mathcal{E}(\rho )] &=\chi _{00}+\chi _{33}+2\mathrm{Re}(\chi
_{03})\langle Z^{A}\rangle , \\
\Tr [P_{-1}\mathcal{E}(\rho )] &=\chi _{11}+\chi _{22}+2\mathrm{Im}(\chi
_{12})\langle Z^{A}\rangle ,\label{eq:stab-eqs}
\end{split}
\end{equation}
where $P_{+1}=P_{\Phi ^{+}}+P_{\Phi ^{-}}$, $P_{-1}=P_{\Psi
^{+}}+P_{\Psi ^{-}}$ and $\langle Z^{A}\rangle =\Tr (\rho Z^{A})\neq 0$ because of our
choice of a non-maximally entangled input state. The experimental data, $\Tr[P_{+ 1}\mathcal{E}(\rho )]$ and $\Tr[P_{- 1}\mathcal{E}(\rho )]$, are exactly the probabilities of no bit-flip error and a bit-flip error on
qubit $A$, respectively. Since we already know the $\{\chi _{mm}\}$ from the
population measurement, we can determine $\mathrm{Re}(\chi _{03})$ and $%
\mathrm{Im}(\chi _{12})$. After measuring $Z^{A}Z^{B}$ the system is in
either of the states $\rho _{\pm 1}=P_{\pm 1}\mathcal{E}(\rho )P_{\pm 1}/%
\Tr [P_{\pm 1}\mathcal{E}(\rho )]$. Next we measure the expectation
value of an element of the normalizer $N(\mathcal{S})$ of the stabilizer group $\mathcal{S}$: 
$\mathcal{N}(\mathcal{S})=\{ g\in P_n: [\mathcal{S},g]=0\}$ (the normalizer preserves the stabilizer under conjugation). For example, we can measure $N=X^{A}X^{B}$, and obtain the measurement results
\begin{equation}
\label{eq:norm-eqs}
\begin{split}
\Tr [N\rho _{+1}]&=\frac{(\chi _{00}-\chi _{33})\langle N\rangle +2i\mathrm{Im}(\chi _{03})\langle Z^{A}N\rangle}{\Tr [P_{+1}\mathcal{E}(\rho )]}\\
\Tr [N\rho _{-1}]&=\frac{(\chi _{11}-\chi _{22})\langle N\rangle -2i\mathrm{Re}(\chi _{12})\langle Z^{A}N\rangle }{\Tr [P_{-1}\mathcal{E}(\rho )]},
\end{split}
\end{equation}
where $\langle Z^{A}\rangle $, $\langle N\rangle $, and $\langle
Z^{A}N\rangle $ are all non-zero and already known. \cref{eq:stab-eqs,eq:norm-eqs} provide a linear system that can be straightforwardly inverted to extract the four independent
real parameters needed to calculate the coherence elements $\chi _{03}$ and $%
\chi _{12}$. It is simple to verify that a simultaneous measurement of the
stabilizer and normalizer elements $Z^{A}Z^{B}$ and $X^{A}X^{B}$, is again a BSM. 

In order to complete the measurement of the remaining coherence elements, we perform an appropriate basis change by preparing the input states $H^{A}H^{B}|\Phi _{\alpha }^{+}\rangle $
and $S^{A}S^{B}H^{A}H^{B}|\Phi _{\alpha }^{+}\rangle $, which are the
eigenvectors of the stabilizer operators $X^{A}X^{B}$ and $Y^{A}Y^{B}$. Here
$H$ and $S$ are single-qubit Hadamard and phase gates acting on the systems $%
A$ and $B.$ At the output, we measure the stabilizers and a corresponding
normalizer element, e.g., $Z^{A}Z^{B}$, which are again equivalent to a standard
BSM, and can be expressed by measuring the Hermitian operators $P_{\Phi
^{+}}\pm P_{\Psi ^{+}}$ and $P_{\Phi ^{-}}\pm P_{\Psi ^{-}}$ (for the input
state $H^{A}H^{B}|\Phi _{\alpha }^{+}\rangle )$, and $P_{\Phi ^{+}}\pm
P_{\Psi ^{-}}$ and $P_{\Phi ^{-}}\pm $ $P_{\Psi ^{+}}$ (for the input state $%
S^{A}S^{B}H^{A}H^{B}|\Phi _{\alpha }^{+}\rangle $). Figure~\ref{fig:QPT}(c)
illustrates the DCQD scheme.

Overall, in DCQD we only need a single fixed measurement apparatus capable
of performing a Bell state measurement, for a complete characterization of
the dynamics. This measurement apparatus is used in four ensemble
measurements each corresponding to a different input state. Figure~\ref%
{fig:QPT}(c) and \cref{dcqd-tab} summarize the preparations required for
DCQD in the single qubit case. This table implies that the required
resources in DCQD are as follows: (a) preparation of a maximally entangled
state (for population characterization), (b) preparation of three other known
(non-maximally) entangled states (for coherence characterization), and (c) a
fixed Bell-state analyzer.

We have assumed no SPAM errors. However, this assumption can be relaxed in certain situations, in particular when the imperfections are already known~\cite{Nigg:2013aa,Graham:2013aa}. 

The DCQD scheme has been generalized to higher-dimensional
quantum systems, which includes the multi-qubit case~\cite{MohseniLidar:07}. The primary advantage of DCQD over all separable QPT methods (in particular, SQPT and separable AAPT; recall that non-separable AAPT requires many-body interactions, which are experimentally unrealizable) is that the minimal number of required experimental configurations is reduced quadratically from $D^{4n}$ to $D^{2n}$, for $n$ qudits of dimension $D$ each. For example, the number of experimental configurations required for systems of three or four physical qubits is
reduced from $\sim 5 \times 10^3$ and $\sim 6.5 \times 10^4$ in SQPT to $64$ and
$256$ in DCQD, respectively~\cite{Mohseni:2008ly}. Moreover, only a single ensemble measurement is required for a commuting set of observables, as was recently rediscovered in the context of entanglement-enhanced learning scheme of Pauli channels\cite{seif2024entanglementenhancedlearningquantumprocesses}.

\subsection{Gate Set Tomography}

The main disadvantage of QPT is its sensitivity to SPAM errors. Building on quantum state and process tomography, gate set tomography (GST) was introduced for estimation of gate performance in quantum processors while avoiding the impact of SPAM errors, specifically calibration of gate errors and measurements~\cite{ibm_gst_2013, sandia_gst_2013, sandia_gst_2014}. Since its introduction, there have been many experimental implementations that have assessed gate errors by utilizing GST~\cite{Blume_2017, Rol_2017, Mavadia_2018, Hempel_2018, Song_2019, Cincio_2021,  Hong_2020, Joshi_2020, Zhang_2020, Ware_2021}.

Unlike quantum process tomography, GST estimates the performance of a gate set without relying on prior calibration of a system, making GST a more generic method for gate performance characterization. Like QPT, GST differs from RB in that it provides a complete characterization of gates, whereas RB provides a single parameter. As an example, RB is insensitive to coherent errors, whereas GST attempts to amplify the effect of coherent gate errors such as rotation and phase errors~\cite{Dehollain_2016}.

GST is used for estimating the performance of the reconstructed gates after applying the gates defined in a gate set, as we describe below. To start the derivation of the GST method, we need to consider the probability of measuring the system along a certain axis using a measurement operator $M_i$ from a POVM (see \cref{sec:QDM}) performed after an arbitrary single gate $G$ is applied to a system in the state $\rho$, which can be written as:

\begin{equation}
\label{eq:gst_prob}
p_i = \superbra{M_i}G\superket{\rho} = \Tr (M_i G \rho),
\end{equation}
where $\superket{\cdot}$ ($\superbra{\cdot}$) represents the super-ket (bra) operators in Hilbert-Schmidt space. 

The superoperator notation is commonly used to simplify manipulations with quantum maps \cite{Kitaev2002}. Using this formalism density matrices $\rho$ acting on a Hilbert space of dimension $D$ are ``vectorized'' and denoted as $\superket{\rho}$, a vector in a Hilbert-Schmidt space of dimension $D^2$. The superket $\superket{\rho}$ is a vector created from the matrix representation of an operator by consecutively stacking its columns. Similarly, quantum operations are represented as matrices of dimension of $D^2 \times D^2$ (in the same basis as $\superket{\rho}$) with the inner product defined as $\langle{\langle{A}}\superket{B}= \Tr{(A^{\dagger}B)}$, where $A$ and $B$ can be density operators. 

Assuming a processor can perform $N_\rho$ state preparation gates and $N_G$ distinct gates with $N_M$ measurement operators, where the $i$'th measurement can result in $N^{(i)}_{M}$ distinct outcomes, we can write down the gate set of the system as  

\begin{equation}
\label{eq:gst_set}
\mathbb{G} = \left \{ \{\superket{\rho^{(j)}}\}^{N_{\rho}}_{j=1}; \{G_j\}^{N_G}_{j=1}; \superbra{M^{(m)}_{j}}^{ N^{(j)}_{M}, N_M}_{m=1,j=1} \right \}.
\end{equation}
This gate set consists of all possible operations that one can perform on the system, where all the gates are dependent on each other. \cref{eq:gst_prob} implies that the measurement probability for any combination of measurement, gate, and initial state is invariant under a similarity transformation $S$, i.e., 
\begin{equation}
\begin{split}
    \superbra{M_j} &\rightarrow \superbra{M_j} S^{-1}, \\
    \superket{\rho^{(j)}} &\rightarrow S \superket{\rho^{(j)}}, \\
    G &\rightarrow S G S^{-1}.
\end{split}
\end{equation}
The outcome of a gate set $\mathbb{G}$ remains invariant up to an arbitrary invertible $S$, which suggests choosing a ``gauge'' $S$ that results in the highest fidelity after reconstructing the gate set.

\begin{figure}[t]
\centering
\includegraphics[width=0.47\textwidth]{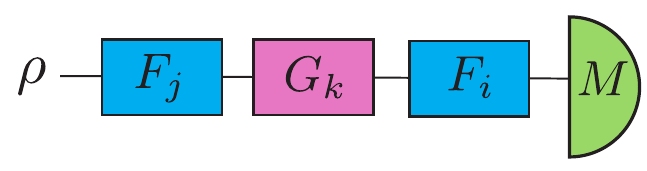}
\caption{Gate set tomography circuit consisting of the fiducial gates $F_{i}$ ($F_{j}$) as well as the measurement operators $M$ and the native gate $G_{k}$ performed on the initial state $\rho$. Figure adapted from Ref.~\citenum{Nielsen_2021}. Copyright 2021 Quantum.}
\label{fig:gst_circuit}
\end{figure}

To make GST independent of the prior calibration of the system and make it robust against SPAM errors, we need to introduce fiducial gates and rewrite the state preparation and measurement gates in \cref{eq:gst_set} as

\begin{equation}
\label{eq:F-transf}
\begin{split}
   \superket{\rho^{(j)}} &\rightarrow F_j \superket{\rho^{(j)}}, \\
   \superbra{M_i} &\rightarrow \superbra{M_i} F_i,
\end{split}
\end{equation}

\noindent where $F_i$ is the fiducial circuit constructed using the gates $\{G_j\}$ in the gate set. \cref{fig:gst_circuit} shows the circuit required to perform GST on a system.

The most important caveat here is that one needs to make sure that these fiducial circuits form an informationally-complete set of state preparation and measurement operators, i.e., measurements that allow for a complete reconstruction of the quantum state~\cite{Scott_2006}. 

As an example, we can form a gate set such that $\{G\} = \{ I, \sqrt{X}, \sqrt{Y} \} = \{G_0, G_1, G_2\}$ with fiducials $\{F\} 
= \{ I, \sqrt{X}, \sqrt{Y}, \sqrt{X} \circ \sqrt{X} \}$, where $\sqrt{X}$ and $\sqrt{Y}$ represent a $\pi/2$ rotation about the Bloch sphere $x$ and $y$ axis, respectively. The choice of $\sqrt{X} \circ \sqrt{X} = X$ instead of $X$ is made here to reduce the experimental cost of implementing the protocol. The key point in this example is that we can span the Bloch sphere (informational completeness) after applying the fiducial gates to the states and measurement operators. 

In the final step of GST, we need to define and optimize a likelihood function such that it reproduces the desired gates based on the experimental data. To this end, we first define the outcome probability of applying the GST protocol on the system of interest as~\cite{Greenbaum_2015}

\begin{equation}
\label{eq:p_ijk}
    p_{ijk} (\vec{\theta}) = \superbra{M_i(\vec{\theta})} F_i(\vec{\theta}) G_k(\vec{\theta}) F_j(\vec{\theta}) \superket{\rho(\vec{\theta})},
\end{equation}
where we combined \cref{eq:gst_prob,eq:F-transf}. Here $\vec{\theta}$ is a vector consisting of a set of parameters that need to be optimized using the maximum likelihood estimation (MLE) method, i.e., the elements of the $T$ matrix from the Cholesky decomposition introduced below.

The likelihood function can be simplified and written in a normalized form as

\begin{equation}
\label{eq:gst_mle}
    \mathcal{L}(\mathbb{G}) = \sum_{ijk} (m_{ijk} - p_{ijk} (\vec{\theta}))^2 / \sigma^2_{ijk},
\end{equation}
with $m_{ijk}$ representing the measured values in the experiment corresponding to the probabilities $p_{ijk}$, and $\sigma^2_{ijk}$ is the variance in the measurement outcome $m_{ijk}$. The remaining task is to minimize the likelihood function [\cref{eq:gst_mle}] using the process matrix $\bm{\chi}$.

We can write \cref{eq:chi-from-E} in terms of the process matrix $\bm{\chi}_G$ for a gate superoperator $\mathcal{G}$ as

\begin{equation}
\label{eq:gst_chi}
    \mathcal{G}(\rho) = \sum_{i,j=0}^{D^2-1} (\bm{\chi}_G)_{ij} E_i \rho E_j ,
\end{equation}
where $\{E_i\}$ is a Hermitian operator basis for $\mathcal{B}(\mathcal{H})$.
Since $\bm{\chi}_G$ is positive semi-definite we can perform a Cholesky decomposition and rewrite it as $\bm{\chi}_G = T^{\dagger} T$ where $T$ is a lower triangular matrix composed of complex elements on the off-diagonal and real elements on the diagonal \cite{Householder_2013}. 

The trace-preservation property of the density matrix $\Tr (\mathcal{G}(\rho)) = \Tr (\rho)$ leads to a completeness relation that can be used to reduce the number of free parameters in $\bm{\chi}_G$ through $D^2$ linear equalities:

\begin{equation}
\label{eq:gst_comprel}
    \sum_{i,j=0}^{D^2-1} \chi_{ij} \Tr (E_i E_k E_j) = \delta_{k0} \ , \quad k = 0,\dots,D^2-1.
\end{equation}
As mentioned above, $\vec{\theta}$ is composed of the matrix elements of $T$, now seen to be $(D^4-D^2)/2$.

Rewriting the likelihood function in terms of the process matrices using \cref{eq:gst_chi} we obtain

\begin{equation}
\label{eq:gst_mle_chi}
\begin{split}
    p_{ijk}(\vec{\theta}) &= \superbra{M_i(\vec{\theta})} F_i(\vec{\theta}) G_k(\vec{\theta}) F_j(\vec{\theta}) \superket{\rho(\vec{\theta})} \\
    &= \Tr [M_i F_i\{G_k(F_j\rho)\}] \\ 
    &= \sum_{mnrstu} (\bm{\chi}_{F_j})_{tu} (\bm{\chi}_{G_k})_{rs} (\bm{\chi}_{F_i})_{mn} \times  \\
    &\qquad \Tr (M_i E_t E_r E_m \rho E_n E_s E_u).
\end{split}
\end{equation}
By substituting this expression into \cref{eq:gst_mle}, observing that since $M_i$ is a POVM element we have $I - M_i \geq 0$, along with the completeness relation [\cref{eq:gst_comprel}] and the unit-trace condition of the density matrix $\Tr (\rho) = 1$, one can solve the MLE problem by varying the elements of the vector $\vec{\theta}$ (elements of the $T$ matrix). Note that as should be clear from the example we discussed above, the process matrices $\bm{\chi}_G$ and $\bm{\chi}_F$ are related and need to be defined in terms of each other. Moreover, the measurement $M_i$ and state $\rho$ also need to be parameterized similarly to $\bm{\chi}_G$ in terms of their Cholesky decompositions. Optimization of the parameters of $\vec{\theta}$ after minimizing the likelihood function [\cref{eq:gst_mle}], allows us to reconstruct the entire gate set as well as the process matrices $\bm{\chi}_G$ and $\bm{\chi}_F$ using the Cholesky decomposition mentioned above.

In summary, the following steps are needed to implement the GST protocol: 
\begin{enumerate}
    \item Initialize the qubit in a pure state: $\rho = \ketb{0}{0}$.
    \item For a particular choice of $i,j,$ and $k$ apply the gate sequence $F_i G_k F_j$.
    \item Measure the POVM $M_i$. 
    Typically in experiments, $M_i = \ketb{0}{0}$.
    \item Repeat steps 1-3 many times (typically ${\sim}$ 2000). Average and record the results to obtain $m_{ijk}$. 
    \item Repeat steps 1-4 for all $i,j$, and $k$.
    \item Solve the MLE problem using the likelihood function defined in \cref{eq:gst_mle} to extract the optimized vector $\vec{\theta}$ to reconstruct the entire gate $\mathbb{G}$ set, as well as the process matrices using the aforementioned Cholesky decomposition. 
\end{enumerate}

Another variant of this protocol, known as long-sequence GST, has been developed to amplify and highlight the source of errors \cite{Nielsen_2021}. The main difference in this method is that we apply the gates $G_k$ multiple times to amplify the role of coherent errors. The sequence applied is $F_i (G_k)^l F_j$ and $l$ can be varied to estimate the error rates caused by the miscalibration in gates. 

In summary, GST provides a comprehensive measure of gate performance in quantum processors. It gives all the details one needs to study and calibrate gates. However, despite recent efforts to simplify the implementation of GST in experiments \cite{Nielsen_2020}, GST has major drawbacks such as its inability to capture non-Markovian noise \cite{Proctor2020} and being extremely resource intensive. For example, to reconstruct the complete gate set for a single qubit one needs to conduct 80 distinct experiments with various gate sequences, with this number increasing to over 4,000 for 2-qubit gates \cite{Greenbaum_2015}. Note that, as mentioned above, each experiment needs to be averaged over $\sim 2000$ runs to minimize the sampling error, making the GST protocol even more resource-intensive. This can make GST impractical in actual experiments, a consequence of the over-completeness of this method. Recent work has shown that the number of experiments in the two-qubit case can be reduced by a factor of $\sim$ 8 to elicit only the right amount of information required for gate calibration \cite{Rudinger_2023}.

\subsection{Direct Fidelity Estimation}
\label{sec:DFE}

Direct fidelity estimation (DFE)\cite{da_Silva_Landon_Cardinal_Poulin_2011,Flammia_Liu_2011} can be used for estimating how close an implementation of a unitary gate is to the ideal gate, via the average output fidelity metric for quantum maps.
That is, DFE provides a single number between 0 and 1 to judge the given gate implementation, where 1 represents the ideal gate implementation.
The advantages of this method are that it does not make any assumptions about the underlying noise model and is relatively inexpensive and easy to implement.
However, one crucial drawback is that DFE assumes negligible SPAM errors.

\begin{figure}[t]
\centering
\includegraphics[width=0.4\textwidth]{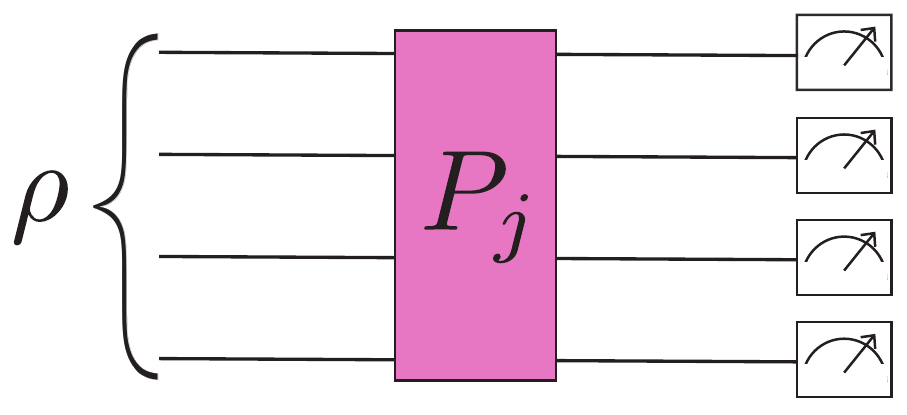}
\caption{Schematic of DFE: To estimate the state fidelity, expectations of random Pauli strings $P_j$ are calculated for the input state $\rho$. The distribution of the random Pauli strings depends on the state with respect to which the fidelity is being calculated.}
\label{fig:dfe_circuit}
\end{figure}

DFE essentially provides a Monte Carlo technique to estimate the Uhlmann fidelity between a given and a target \emph{pure} state.
If the given state is $\rho\in\mcBH$ and the target pure state is $\sigma\in\mcBH$, then the Uhlmann fidelity [recall \cref{eq:F-process}] reduces to
\begin{equation}
F(\sigma,\rho) = \Tr{(\sigma \rho)}.
\end{equation}
Let $\{ P_k\}$ denote the set of Pauli group operators (including identity) in $\mcBH$, normalized so that $\Tr( P_k P_l) = D \delta_{kl}$, where $d = \dim(\mc{H})$. We can then expand 
\begin{equation}
\sigma = \frac{1}{D}\sum_k \sigma_k  P_k\ , \quad 
\sigma_k = \Tr(\sigma  P_k),
\end{equation} 
and similarly for $\rho$. Next, we note that we can recast the fidelity as an expectation over an appropriately defined random variable $X$ that can take values
\begin{equation}
x_k = \frac{\rho_k}{\sigma_k} 
\end{equation}
with probability
\begin{equation}
p_k = \mathbb{P}(X=x_k) = \frac{\sigma_k^2}{D} .
\label{eq:dfe-normalized-prob}
\end{equation}
To see this, note that 
\begin{equation}
\begin{split}
\Tr{(\sigma \rho)} &= \frac{1}{D^2}\sum_{k,l}\sigma_k \rho_l \Tr( P_k  P_l) \\
&= \frac{1}{D} \sum_k \sigma_k \rho_k = \sum_k p_k x_k = \< X\> .
\label{eq:fidelity-with-pauli-coeffs}
\end{split}
\end{equation}
Note that \cref{eq:dfe-normalized-prob} denotes a normalized probability distribution.
The expectation $\< X\>$ can then be approximated using Monte Carlo techniques, assuming that we are able to sample from the distribution $\mathbb{P}$.
The rigorous bound on the error of the Monte Carlo procedure is given by the following result (adapted from Theorem 1 of Ref.~\citenum{Flammia_Liu_2011}):
\begin{mytheorem}
\label{th:F-DFE-estimate}
Given that
\begin{itemize}
\item $\{i_1, \dotsc, i_N\}$ are $N$ indices sampled from the probability distribution $\mathbb{P}$
\item each $\Tr(\rho  P_{i_k})$ is estimated using $N_2^{[k]}$ samples
\end{itemize}
the fidelity estimate $\bar{F}$ can be obtained within an error margin of the true fidelity $F(\sigma,\rho)$ given by

\begin{equation}
\begin{split}
\mathbb{P}( &| F-\bar{F} | <\epsilon_1 + \epsilon_2) \leq \frac{1}{N\epsilon_1^2} + \\
& 2 \exp{\left[- \frac{\epsilon_2^2 N^2}{2} \left( \sum_{k=1}^{N}\frac{1}{\sigma_{i_k}^2 N_2^{[k]}} \right)^{-1} \right]}
\end{split}
\end{equation}

where $\sigma_i = \Tr(\sigma  P_i)$. Here $\epsilon_1$ and $\epsilon_2$ are the errors associated with the Monte Carlo estimate and experimental estimate of $\Tr(\rho  P_{i_k})$ respectively.
\end{mytheorem}

As in \cref{sec:QDM,AAPT-sec}, we can invoke the Choi-Jamiołkowski isomorphism\cite{Jamiolkowski:72,Choi:75} between quantum processes and quantum states. We can then extend DFE to benchmarking gates. As in \cref{eq:Choi-state}, define the Choi states $\rho_{\mathcal{U}} = (\mathbb{I}\otimes\mathcal{U}) (\ketb{\phi}{\phi})$ and $\rho_{\tilde{\mathcal{U}}} = (\mathbb{I}\otimes\tilde{\mathcal{U}}) (\ketb{\phi}{\phi})$, where $\mathcal{U}$ represents the ideal unitary gate operation and $\tilde{\mathcal{U}}$ represents the actual (noisy) experimental implementation.
Then the benchmark metric that is calculable via DFE is the average output fidelity $\bar{F}(\mathcal{U}, \tilde{\mathcal{U}})$, defined as the average of the process fidelities [\cref{eq:F-process}] of $\mathcal{U}$ and $\tilde{\mathcal{U}}$ over all possible pure states chosen uniformly at random.
It can be shown that\cite{Horodecki:1999ab}
\begin{equation}
\bar{F}(\mathcal{U}, \tilde{\mathcal{U}}) = \frac{D F_{\mathrm{proc}}(\rho_{{\mathcal{U}}}, \rho_{\tilde{\mathcal{U}}})+1}{D+1}
\end{equation}
where the expression on the right is calculable using DFE.
A naive implementation would require constructing the maximally entangled state $\ket{\phi}$. 
However, one can sidestep the issue of preparing $\ket{\phi}$ in experiments by simplifying the expression for $F_{\mathrm{proc}}(\rho_{{\mathcal{U}}}, \rho_{\tilde{\mathcal{U}}})$:
\begin{equation}
\label{eq:F-DFE}
\begin{split}
&F_{\mathrm{proc}}(\rho_{{\mathcal{U}}}, \rho_{\tilde{\mathcal{U}}})
  = \frac{1}{D^2} \sum_{j,k} (\rho_{{\mathcal{U}}})_{jk}  (\rho_{\tilde{\mathcal{U}}})_{jk}\\
  &\quad = \frac{1}{D^4} \sum_{j,k}  \Tr{({\mathcal{U}}( P_j^T)  P_k)} \Tr{(\tilde{\mathcal{U}}( P_j^T)  P_k)}\\
  &\quad = \frac{1}{D^4} \sum_{j,k}  \Tr{({\mathcal{U}}( P_j)  P_k)} \Tr{(\tilde{\mathcal{U}}( P_j)  P_k)},
\end{split}
\end{equation}
where the first line follows from \cref{eq:fidelity-with-pauli-coeffs} with $(\rho_\mathcal{E})_{jk} = \Tr(\rho_\mathcal{E}  P_j \otimes  P_k)$, the second line due to the identity $\Tr[(O \otimes\mathcal{E}) (\ketb{\phi}{\phi})] = \frac{1}{D}\Tr(\mathcal{E}(O^T))$), and the third line from the property of Pauli matrices: $ P^T = \pm  P^T$. 
The expression then no longer involves the entangled state $\ket{\phi}$.
The corresponding experimental procedure becomes:
\begin{enumerate}
\item Determine the tuple of Pauli operators $( P_j,  P_k)$ for which $\Tr{({\mathcal{U}}( P_j)  P_k)} \neq 0$. Call this set $A$.
\item Sample a tuple of Pauli operators, each of length $\log_2(D)$, from the set $A$ according to the probability distribution $\mathbb{P}( P_j,  P_k) = \Tr{(\mathcal{U}( P_j) P_k)^2}/D^4$. This is done via importance sampling. Note that here it is relatively easy to sample because it is possible to calculate the entire probability distribution.
\item Prepare the $D$ eigenstates for $ P_j$, which require only local operations.
\begin{enumerate}
\item For each eigenstate, apply $\tilde{\mathcal{U}}$, and calculate the expectation of $ P_k$ (which also requires only local operations and measurements). Call the estimate for the expectation $\tilde{a_i}$, and the eigenvalue of the eigenstate $\lambda_i$.
\item The estimate for $\Tr{(\tilde{\mathcal{U}}( P_j)  P_k)}$ is then $\sum_{i} \lambda_i \tilde{a_i}$. Store $\frac{\Tr{(\tilde{\mathcal{U}}( P_j)  P_k)}}{\Tr{({\mathcal{U}}( P_j)  P_k)}}$ as the estimate for the sample.
\end{enumerate}
\item Repeat the procedure $N$ times. The average of the sample estimates gives $F(\rho_{{\mathcal{U}}}, \rho_{\tilde{\mathcal{U}}})$.
\end{enumerate}
Alternatively, DFE can also be used for calculating the entanglement fidelity given by $F_e = \Tr(\mathcal{U}^\dagger \tilde{\mathcal{U}})/D^2$ where $\mathcal{U}$ and $\tilde{\mathcal{U}}$ are both now treated as matrices acting via left multiplication to give the output state under the corresponding channel\cite{Flammia_Liu_2011}.

Theoretically, the number of samples required for DFE scales exponentially with the number of qubits $n$ in the worst case. However, for most states of interest, the number of samples required is only polynomial in $n$. In practice, DFE is efficient: a variant of it was demonstrated on 7 qubits \cite{Lu_Li_Trottier_Li_Brodutch_Krismanich_Ghavami_Dmitrienko_Long_Baugh_2015}, while vanilla DFE has been used for a benchmarking implementation of a Toffoli gate using superconducting qubits \cite{Fedorov_Steffen_Baur_da_Silva_Wallraff_2012}. However, DFE has been limited in its adaptation to benchmarking due to its susceptibility to SPAM errors; in this sense, alternatives such as RB (\cref{sec:rb}) that are robust to SPAM errors are preferred~\cite{Wallman_Flammia_2014,Erhard_Wallman_Postler_Meth_Stricker_Martinez_Schindler_Monz_Emerson_Blatt_2019,Roth_Kueng_Kimmel_Liu_Gross_Eisert_Kliesch_2018}.

DFE has been extended to simpler and more efficient measurement schemes for fidelity estimation, and tested on multi-qubit entangled states on both trapped-ion\cite{Kalev_Kyrillidis_Linke_2019,Seshadri_Ringbauer_Spainhour_Monz_Becker_2024,Seshadri_Ringbauer_Spainhour_Blatt_Monz_Becker_2024} and superconducting quantum devices\cite{Kalev_Kyrillidis_Linke_2019}. Other uses of DFE include tuning control for logic gates in error correcting codes \cite{kosut_adaptive_2013}, and calculating minimum gate fidelities \cite{Lu_Sim_Suzuki_Englert_Ng_2020}.

\subsection{Process Fidelity Estimation}
\label{sec:process-fid}

An alternative and generalization of DFE is the Process Fidelity Estimation (PFE) method introduced in Ref.~\citenum{baumer2024quantum}. This method is also concerned with estimating the process fidelity between an ideal and actual, noisy process, but generalizes DFE to also account for measurements. As such, it is applicable to entire quantum algorithms.

Consider a projective measurement in the $\sigma_z$-basis described by $\Pi (\rho) = \sum_k \Pi_k(\rho) = \sum_k \pi_k \rho \pi_k$ (where $\pi_k = \ketb{k}{k}$ and $\{\pi_k\}$ are a complete set of orthogonal projectors), an ideal unitary map $\mathcal{U}(\rho)=U\rho U^\dag$, and its noisy implementation $\tilde{\mathcal{U}}$. Let the ideal map be $\Phi=\Pi  \cU$ and its noisy version be $\tilde{\Phi} = \Pi  \tilde{\mathcal{U}}$. We can always formally decompose the noisy unitary map as $\tilde{\mathcal{U}} = \Lambda_U \cU$ where $\Lambda_U = \tilde{\mathcal{U}}\cU^\dag$ is a $U$-dependent noise map that captures all the noise acting during the noisy implementation of the ideal unitary. 

Here, the orthonormal basis states $\{\ket{i}\}$ in the maximally entangled state $\ket{\phi} = \frac{1}{\sqrt{D}}\sum_{i=1}^D \ket{i}\otimes\ket{i}$ used to construct the Choi state [\cref{eq:Choi-state}] can be chosen to be the eigenstates of the projectors $\{\pi_k\}$, i.e., $\pi_k\ket{i} = \delta_{ki}\ket{k}$.
Since $\Phi\cU^\dag =\Pi$ and $\tilde{\Phi}\cU^\dag =\Pi\Lambda_U \equiv \tilde{\Pi}$, and the Uhlmann fidelity is unitarily invariant, we have
\begin{equation}
\label{eq:F-Uinv}
F_{\mathrm{proc}}(\rho_{\Phi},\rho_{\tilde{\Phi}}) = F_{\mathrm{proc}}(\rP,\rtP)\ .
\end{equation}
Note that 
\begin{equation}
\begin{split}
\Pi(A) &= \sum_k \pi_k A \pi_k =  \sum_k \bra{k}\! A\ket{k}\pi_k \ .
\end{split}
\end{equation}
Writing $e_{ij}\equiv\ketb{i}{j}$, the Choi state of $\tP$ is given by
\begin{equation}
\begin{split}
    \rtP &\equiv (\mathbb{I} \otimes \tP)(\ketb{\phi}{\phi})\\
    &=\frac{1}{D}\sum_{i,j=1}^D e_{ij}\otimes \Pi[\Lambda_U(e_{ij}) ]\\
    &=\frac{1}{D}\sum_{k=1}^D \sum_{i,j=1}^D\bra{k}\Lambda_U(e_{ij})] \ket{k}e_{ij}\otimes\pi_k\\
    &=\frac{1}{D}\sum_k \tsk\otimes\pi_k\ , 
\end{split}
\end{equation}
with
\begin{equation}
    \tsk \equiv \sum_{ij} \bra{k}\Lambda_U(e_{ij}) \ket{k}e_{ij}\ . 
\end{equation}
Similarly, in the noiseless case, i.e., after dropping the tilde everywhere:
\begin{equation}
\label{eq:rcVold}
\rP \equiv (\mathbb{I} \otimes \Pi) (\ketb{\phi}{\phi}) = \frac{1}{D}\sum_k\sk\otimes\pi_k\ ,
\end{equation}
where:
\begin{equation}
    \sk \equiv \sum_{i,j} \bra{k}e_{ij}\ket{k}e_{ij} = \sum_{i,j} \delta_{ki}\delta_{jk}e_{ij} =\pi_k \ ,
    \label{eq:sk=pik}
\end{equation}
so that 
\begin{equation}
\label{eq:rcV}
\rP = \frac{1}{D}\sum_k P_k\ , \quad P_k = \pi_k\otimes\pi_k\ , 
\end{equation}
and $\{P_k\}$ is another set of orthogonal projectors: $P_k P_l = \delta_{kl} P_k$. Therefore
\begin{equation}
\label{eq:rhoV^2}
    \rP^2 =  \frac{1}{D^2}\sum_{k,l=1}^D P_k P_l =  \frac{1}{D} \rP ,
\end{equation}
so that $\sqrt{D}\rP$ is the unique positive semi-definite square-root of $\rP$, i.e., $\sqrt{\rho} = \sqrt{D}\rP$.

Thus, combining \cref{eq:F-process,eq:F-Uinv}
\begin{equation}
\begin{split}
F_{\mathrm{proc}}(\Phi,\tilde{\Phi}) &= F_{\mathrm{proc}}(\Pi,\tP) = 
    F(\rP,\rtP) \\ 
    &=\Tr\sqrt{\sqrt{\rP}\rtP\sqrt{\rP}} \\
    &= \Tr\sqrt{d\rP\rtP\rP}\\
    &= \frac{1}{D}\Tr\sqrt{\sum_k a_k P_k}\ ,
\end{split}
\end{equation}
where $a_k = \Tr(\pi_k\tsk)$. Noting that 
\begin{equation}
(\sum_k \sqrt{a_k}P_k)^2 = \sum_{kl} \sqrt{a_k}\sqrt{a_l}P_k\delta_{kl} = \sum_k a_k P_k \ ,
\end{equation} 
we can write
\begin{equation}
    \label{eq:A13}
    F_{\mathrm{proc}}(\Phi,\tilde{\Phi}) = \frac{1}{D}\Tr\sum_k \sqrt{a_k}P_k  = \frac{1}{D} \sum_k \sqrt{a_k} \ .
\end{equation}

Now, since
\begin{equation}
\begin{split}
    a_k &= \bra{k}\tsk\ket{k} 
    =\sum_{i,j} \delta_{ki}\delta_{jk}\bra{k}\Lambda_U(e_{ij})\ket{k}\\
    &= \bra{k}\Lambda_U(\pi_k)\ket{k}\ ,
\end{split}
\end{equation}
this finally yields:
\begin{equation}
\label{eq:F-PFE}    
\begin{split}
    F_{\mathrm{proc}}(\Phi,\tilde{\Phi})
    &=\frac{1}{D}\sum_{k=1}^D\sqrt{\bra{k} \Lambda_U(\pi_k)\ket{k}}\\
    &= \frac{1}{D}\sum_{k=1}^D\sqrt{ \text{Pr}\left(k| \Lambda_U(\pi_k)\right)}\ ,
\end{split}
\end{equation}
where $\text{Pr}(x|\rho)$ is the probability of measuring the output $x$ given a state $\rho$. This result is analogous to the process fidelity expression associated with the DFE method, i.e., \cref{eq:F-DFE}. However, the two methods clearly result in very different operational interpretations. 

Since $\Lambda_U(\pi_k) = \tilde{\mathcal{U}}[\cU^\dag(\pi_k)]$, \cref{eq:F-PFE} has the following operational interpretation: it means that we can determine the fidelity of a noisy unitary that is followed by a measurement, by preparing the time-reversed initial states $\ket{\psi_k} = U^\dag \ket{k}$,  
applying the noisy unitary $\tilde{\mathcal{U}}$ and combining the probabilities to measure the respective outputs $k$. However, this requires the ability to prepare the initial states $\ket{\psi_k}$ with high fidelity, which is a potential drawback of the PFE method. In certain cases this only involves accurately implementable single qubit gates~\cite{baumer2024quantum} (Hadamard and virtual phase gates).

As the number of different initial states $\ket{k}$ increases exponentially, we can sample $m$ different values $\{k_l\}_{l=1..m}$, where $k_l \in \{1, .., 2^n\}$, by drawing randomly from the uniform distribution over all $2^n$ values. 
Note that using the square of the sample average (denoted by an overline) as an estimator for $E[\sqrt{X}]^2$ has an $O(1/m)$ bias.  This can be seen by noting that $E[(\overline{\sqrt{X}})^2] = E[(\frac{1}{m}\sum_i\sqrt{X_i})^2] = E[\frac{1}{m^2}(\sum_{i\neq j} \sqrt{X_iX_j}+ \sum_i X_i] = \frac{m-1}{m}E[\sqrt{X}]^2 + \frac{1}{m}E[X]$. However, we can construct an unbiased estimator of $E[\sqrt{X}]^2$ by using $\frac{m}{m-1} (\overline{\sqrt{X}})^2 - \frac{1}{m-1} \overline{X}$, and estimate the process fidelity via 
\begin{equation}
\label{eq:F-PFE-estimate}
    \begin{split}
    &\left[F_{\mathrm{proc}}(\Phi,\tilde{\Phi})\right]^2 = \left[\frac{1}{D}\sum_{k=1}^{2^n}\sqrt{ \text{Pr}\left(k| \Lambda_U(\pi_k)\right)}\right]^2\\
    &\quad \approx \frac{m}{m-1} \left[\frac{1}{m}\sum_{l=1}^m\sqrt{ \text{Pr}\left(k_l|  \Lambda_U(\pi_k)\right)}\right]^2  \\
    &\qquad - \frac{1}{m(m-1)}\sum_{l=1}^m \text{Pr}\left(k_l|  \Lambda_U(\pi_k)
    )\right).
\end{split}
\end{equation}
where $l$ runs over the samples.

This method was used recently in an experimental demonstration of the quantum Fourier transform using mid-circuit measurements~\cite{baumer2024quantum}. Given its novelty, it has not yet been implemented in other experiments.

\subsection{Cross-Entropy Benchmarking}
\label{sec:xeb}

Cross-entropy benchmarking (XEB) is a scalable noise characterization and benchmarking scheme that estimates the fidelity $F$ of a $n$-qubit circuit composed of single- and two-qubit gates. The procedure is based on sampling bitstrings from the output probability distribution of a quantum circuit of interest and was first proposed to characterize the fidelity of running random quantum circuits on a noisy quantum processor, a task known as random circuit sampling (RCS)\cite{neill_blueprint_2018}. Later, it was used to produce indirect evidence of quantum supremacy~\cite{arute_quantum_2019}. Following these developments, a new gate characterization method called \emph{RCS benchmarking} was built on top of XEB and RCS, where the overall noise strength of a circuit is estimated using XEB of random circuits~\cite{liu_benchmarking_2022}.

Consider the task of sampling bitstrings from the quantum state generated by a circuit. An ideal circuit $C_U$, which implements a unitary $U$ prepares an ideal pure state $\rho_U= \ketb{\psi_U}{\psi_U}$, which can be used for sampling. In reality, the bitstrings are drawn from a noisy state $\rho_{\tilde{U}}$ prepared by the noisy circuit $C_{\tilde{U}}$ instead. We can model the effect of noise in terms of a simple CPTP map:
\begin{equation}
\label{eq:xeb_state}
    \rho_{\tilde{U}} = F_U\ketb{\psi_U}{\psi_U} + (1-F_U)\sigma_U,
\end{equation}
where $\sigma_U$ denotes an ``error state'' that captures the effect of the noise, and $F_U = \bra{\psi_U}\rho_{\tilde{U}}\ket{\psi_U}$ is the fidelity we would like to estimate. Methods such as DFE (see \cref{sec:DFE}) satisfy precisely this task but have poor worst-case scaling, requiring an exponential number of samples in the number of qubits $n$. Instead, to compute $F$, we first define the cross-entropy metric~\cite{elements_of_info_th_covey} between the ideal and noisy outcome probability distributions:

\begin{equation}
    S(p_{\tilde{U}},p_U) = -\sum_{x\in\{0,1\}^n}
    p_{\tilde{U}}(x) \log[p_U(x)],
\end{equation}
where $p_U(x) = \bra{x}\rho_U\ket{x}$ and $p_{\tilde{U}}(x) = \bra{x}\rho_{\tilde{U}}\ket{x}$ denote the ideal and actual probability distribution of observing the bitstring $x$ from the corresponding ideal and actual states $\rho_U$ and $\rho_{\tilde{U}}$ respectively. Note that $p_U(x)$ is the probability of bitstring $x$ computed classically for the ideal quantum circuit $U$, while $p_{\tilde{U}}(x)$ is obtained from running the actual noisy quantum circuit $\tilde{U}$ using a quantum computer.

When the set of circuits $C$ is random, the expectation of the cross entropy is~\cite{boixo_characterizing_2018}:
\begin{equation}
\begin{split}
    &\mathbb{E}_U[S(p_{\tilde{U}},p_U)] \\
    &\quad= -\mathbb{E}_U \left[\sum_{x} p_{\tilde{U}}(x) \log[p_U(x)] \right] \\
    &\quad= \mathbb{E}_U \left[F_U S(p_U,p_U) + (1-F_U) S(p_{\sigma_U},p_U) \right] \\
\end{split}
\end{equation}
where $p_{\sigma_U} = \bra{x}\sigma_U\ket{x}$ and we substituted \cref{eq:xeb_state} in the third line. Under the assumption that the fidelity depends only on the specific gates applied, and not their order, i.e. $F_U = F$, solving for $F$ yields
\begin{equation}
\label{eq:fidelity-xeb}
    F = \frac{\mathbb{E}_U [S(p_{\sigma_U},p_U)] - \mathbb{E}_U [S(p_{\tilde{U}},p_U)]}{\mathbb{E}_U [S(p_{\sigma_U},p_U)] - \mathbb{E}_U [S(p_U,p_U)]}.
\end{equation}
In a loose sense, this reflects an uncorrelated noise model. We further assume that for sufficiently deep circuits, $\sigma_U$ results in a probability distribution $p_{\sigma_U}$ which under expectation behaves like a uniform distribution, i.e., $\mathbb{E}_U[p_{\sigma_U}]$ is a uniform distribution over bitstrings. This is formalized as follows:
\begin{equation}
\label{eq:xeb_assumption}
    \sum_x \bra{x}\sigma_U \ket{x} \log [p_U(x)]
    = \frac{1}{D}\sum_x \log [p_U (x)] +\epsilon,
\end{equation}
where $D=2^n$ and $\epsilon$ decreases sufficiently fast with the number of qubits $n$. This forms the central assumption in XEB methods. For a depolarizing map as in \cref{eq:rb_infid}, the error state is $\sigma_U=I/D$, and \cref{eq:xeb_assumption} is exact. More generally, the error $\epsilon$ in \cref{eq:xeb_assumption} decreases exponentially with $n$. This can be understood as the ``concentration of measure phenomenon" in high-dimensional spaces in the limit of a large number of qubits $n$~\cite{ledoux_concentration_2001}.

\cref{eq:xeb_assumption} makes the expression $\mathbb{E}_U [S(p_{\sigma_U},p_U)]$ classically calculable. Therefore in \cref{eq:fidelity-xeb}, only the term $\mathbb{E}_U [S(p_{\tilde{U}},p_U)]$ needs to be calculated experimentally; the rest are calculated classically.

In practice, instead of using the logarithm to calculate the standard XEB, a linearized version of XEB is considered. The linear XEB is defined as~\cite{arute_quantum_2019}:
\begin{equation}
\label{eq:S_linear}
S_\mathrm{linear}(p_U,p_{\tilde{U}})
= D \sum_{x\in \{0,1\}^n} p_{\tilde{U}}(x) p_U(x) - 1.
\end{equation}
Intuitively, both the standard and linear XEBs are expectations of functions ($\mathbb{E}_{X \sim p_{\tilde{U}}}[\log{p_U(X)}]$ and $\mathbb{E}_{X \sim p_{\tilde{U}}}[p_U(X)d - 1]$ respectively) which map observed bitstrings with higher ideal probabilities to larger values. When the circuits are noiseless, in expectation, the bitstring samples are drawn from the ideal probability distribution of random circuits, which is called the Porter-Thomas  distribution~\cite{boixo_fourier_2017}, an exponential distribution over a random permutation of bitstrings.
This is the limit where $F=1$. In the opposite limit where the bitstring samples are drawn from the uniform distribution, we have $p_{\tilde{U}}(x)=\frac{1}{D}$ and $F=0$. The same analysis as \cref{eq:fidelity-xeb} follows through with the modified, linear XEB definition.

The above exposition describes the theory of using the XEB to calculate the fidelity $F$. Through heuristic arguments~\cite{arute_quantum_2019,boixo_characterizing_2018,Choi_2023}, the unbiased linear cross entropy is considered an estimator of the fidelity itself.

As in the case of DFE (\cref{th:F-DFE-estimate}) and PFE [\cref{eq:F-PFE-estimate}], directly computing the exponentially large sum involved in \cref{eq:S_linear} is impractical, and instead we resort to a sampling method. To construct an unbiased estimator for the linear cross entropy, consider $M$ bitstring samples $S=\{x_i\}_{i=1}^M$ drawn from the probabilities of the noisy output state of a random circuit, the linear cross-entropy fidelity estimator is given by

\begin{equation}
    \label{eq:xeb_estimator}
    \hat{F}_\mathrm{XEB} = \frac{D}{M}\sum_{i=1}^M p_U(x_i)-1.
\end{equation}
Ref.~\citenum{rinott_statistical_2022} proposed the following form for an unbiased linear XEB fidelity estimator:
\begin{equation}\label{eq:xeb_unbiased}
    \hat{F}_\mathrm{uXEB}=\frac{\frac{D}{M}\sum_{i=1}^Mp_U(x_i)-1}{d\sum_{x\in\{0,1\}^n}p_U(x)^2-1}
\end{equation}
Note that one will need to compute all $2^n$ output probabilities in the denominator of \cref{eq:xeb_unbiased} to compute the unbiased linear XEB fidelity estimator. For random circuits, the denominator in \cref{eq:xeb_unbiased} approaches $1$ in log depth \cite{dalzell_random_2022}, hence the two XEB fidelity estimators, \cref{eq:xeb_estimator} and \cref{eq:xeb_unbiased} give the same result as the depth increases. However, the unbiased fidelity estimator $\hat{F}_\mathrm{uXEB}$ is more accurate for circuits with short depth.

In both estimators for XEB, $\{x_i\}$ is sampled from the noisy circuit, while the ideal circuit is classically simulated to provide the values $p_U(x_i)$. A variety of different classical algorithms have been devised for this task, among which tensor networks are the most efficient\cite{huang2020classical,Zlokapa:2023aa,Kechedzhi:2024aa}.

Thus, computing $F_\mathrm{XEB}$ becomes infeasible for large circuit sizes that do not have tractable classical simulations. This computational hurdle motivates using the XEB measure as a way to certify a quantum computing advantage. More precisely, one can demonstrate a quantum advantage over classical algorithms, for the specific task of estimating the XEB values in the context of the computational task of sampling bitstrings from the probability distributions generated by random circuits. This idea was the basis for the first quantum supremacy experiment~\cite{Arute2019}.

Among other things, the experimental linear XEB was shown to be consistent with a simple uncorrelated noise model\cite{Arute2019}. This fact underlines that sampling from random circuits can be used as a way to benchmark noise in quantum circuits. This is known as RCS benchmarking \cite{liu_benchmarking_2022}.

\begin{figure}[t]
\centering
\includegraphics[width=0.47\textwidth]{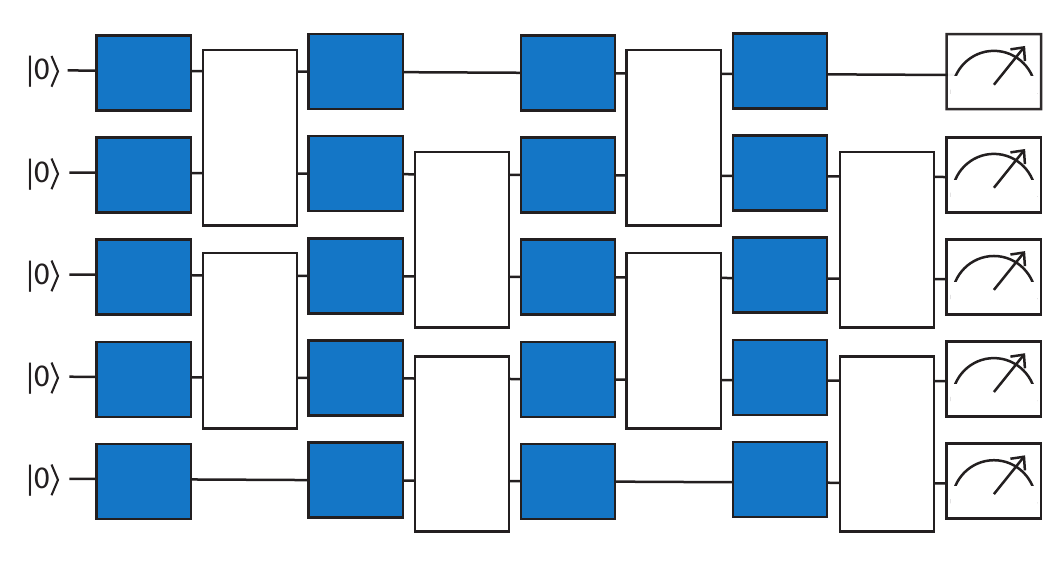}
\caption{RCS benchmarking circuit: layers of 2-qubit gates (white boxes) are interleaved with Haar random 1-qubit gates (blue boxes), followed by measurement in the computational basis at the end. Adapted from Fig.~1 of Ref. \citenum{liu_benchmarking_2022}. Copyright 2021 arXiv.}
\label{fig:rcs_circuit}
\end{figure}

In an RCS benchmarking circuit, similar to the circuit in the IRB protocol (\cref{sec:IRB}), the gate to be characterized is applied in an alternating fashion and interleaved with Haar random single-qubit gates (approximated using unitary $t$-designs\cite{Harrow:2023aa,Haferkamp:2023aa}), followed by measurements in the computational basis; see \cref{fig:rcs_circuit}.
The noise introduced by the imperfect gates is assumed to be described by a Pauli noise model of the form $\mathcal{N}(\rho)=\sum p_\alpha \sigma_\alpha \rho \sigma_\alpha $, where $\sigma_\alpha$ is an $n$-qubit Pauli operator and $p_\alpha$ is the associated error probability.
The RCS benchmark procedure is the following:
\begin{enumerate}
    \item Sample a random circuit $C_i$ from an ensemble of random circuits $\mathrm{RCS}(n,P)$ with $n$ qubits and depth $P$, where $P$ denotes the layers of two-qubit gates independently drawn from the Haar measure on $\mathbb{U}(4)$.
    \item Estimate the fidelity of $C_i$ (denoted $\hat{F}_{P,i}$).
    \item Repeat step 1-2 for $L$ instances of random circuits and obtain the average fidelity $\hat{F}_P=\frac{1}{L}\sum_{i}^L\hat{F}_{P,i}$.
    \item Repeat step 1-3 for varying depths $P \in \{1,\dots, \mathcal{P}\}$ to obtain a sequence of average fidelities $\{\hat{F}_P\}_1^\mathcal{P}$.
    \item Fit the average fidelities $\{\hat{F}_P\}_1^\mathcal{P}$ to an exponentially decaying function $F=A e ^{-\lambda P}$, where the fitted parameter $\lambda$ characterizes the overall noise strength.
\end{enumerate}

The key step in the procedure above is estimating the circuit fidelity $F$, which is defined as the overlap between the output state $\ket{\psi_U}$ of the ideal circuit and the mixed state $\rho_{\tilde{U}}$ generated by an experiment, i.e., $F_U=\bra{\psi_U}\rho_{\tilde{U}} \ket{\psi_U} $. This type of fidelity estimate, in general, requires exponentially many measurements to determine the state $\rho_{\tilde{U}}$. To circumvent this exponential cost, the cross-entropy fidelity estimator in \cref{eq:xeb_estimator} is considered instead. 

Under the assumption of a Pauli noise map, it has been shown, both analytically and numerically~\cite{liu_benchmarking_2022}, that the average fidelities of the random circuits decay exponentially as a function of circuit depth $P$, i.e., $\mathbb{E}[F]=e^{-\lambda P}$ where $\lambda=\sum_\alpha p_\alpha$ , which allows one to extract the effective noise rate of the quantum hardware $\lambda$ by estimating $\mathbb{E}[F]$. Similar to the RB protocol, when fitting the experimentally estimated XEB fidelities $\{\hat{F_P}\}_{P=1}^{P_{\max}}$ to an exponentially decaying function, the depth-independent fitting parameters correspond to SPAM errors, which can be estimated independently. For this reason, the XEB benchmark protocol is considered robust to SPAM errors. The exponential decay characteristic and the related extraction of noise parameters can be generalized beyond Pauli noise maps\cite{Zlokapa:2023aa}.

Another desirable feature of the XEB benchmark is that the fidelity estimation procedure is sample-efficient. To see this, we consider the variance of the unbiased XEB fidelity estimator, which has been shown to be given by \cite{rinott_statistical_2022}:
\begin{equation}
    \operatorname{Var}_{C,S}(\hat{F}_\mathrm{uXEB})=O\Big(\frac{1}{M}+\lambda ^2 (\mathbb{E}[F])^2 \Big).
\end{equation}
This suggests that the fidelity estimator is sample-efficient (compared to, e.g., DFE) and requires $M=O(1/\epsilon^2)$ measurement samples to achieve $\epsilon$ additive accuracy for each instance of random circuits. The first term in the variance expression is the leading term, especially in large-scale experiments, due to the exponential decay of $\mathbb{E}[F]$. In practice, it suffices to collect a large number of samples for a few circuits to estimate $\mathbb{E}[F]$ with a small additive error.

Unlike the standard RB protocol described in \cref{sec:rb}, which involves Clifford gate sequences, the XEB protocol can also be applied to circuits with arbitrary non-Clifford gates. Moreover, RB does not scale easily beyond two or three qubits due to the large circuit depth it requires; XEB, in contrast, is limited only by the classical simulability of the underlying circuit and can extend to up to $\sim 50$ qubits~\cite{liu_benchmarking_2022}. Hence, it can be used to benchmark broader families of quantum gates and quantum circuits. However, the XEB fidelity estimator can be insensitive to certain types of noise. As an example, under a Pauli noise model, local Pauli errors can add up or cancel as they propagate forward in time through random circuits. When the propagated error operators before the measurement are of type $I$ or $Z$, they do not contribute to the XEB fidelity as they do not affect measurements in the computational basis. It has also been shown that the XEB is not a good proxy for the fidelity in the presence of time-correlated errors \cite{gao_limitations_2024}. Overall, the XEB protocol is useful in estimating the total amount of error in a sample-efficient way under certain strong assumptions about the noise. To extract more information about the noise map, alternative noise characterization techniques are needed.

\section{Deterministic Benchmarking}
\label{sec:DB}

\begin{figure}[t]
		\centering
		\includegraphics[width=0.9\columnwidth]{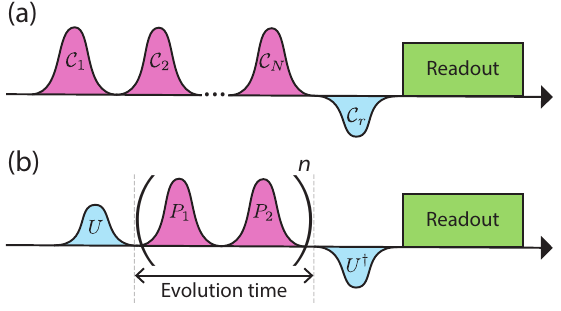}
		\caption{Schematic representations of the gate sequences for (a) randomized benchmarking (RB) and (b) deterministic benchmarking (DB), respectively. RB uses $N$ random Clifford gates $C_i$, followed by a single recovery gate $C_r = C_N^\dag \cdots C_1^\dag$. DB prepares an initial state via $U\ket{0}$, applies $n$ repetitions of a fixed, deterministic two-pulse sequence $P_1 P_2$, and unprepares the initial state via $U^\dag$.}
		\label{fig:dd}
\end{figure}

In this section, we introduce a method we call \emph{deterministic benchmarking} (DB) and validate its utility through experiments conducted on a superconducting transmon qubit \cite{Koch2007}.  Unlike RB, which averages over many and lengthy random Clifford gate sequences to approximate all errors as a depolarizing map and lacks the sensitivity needed to detect small changes in decoherence times and qubit frequencies \cite{Burnett2019, Proctor2020}, DB is highly sensitive to such changes.  As illustrated in \cref{fig:dd}(a,b), DB utilizes a carefully chosen small set of gate-pair sequences that specifically target the real noise affecting single-qubit gates.  The DB protocol is significantly more efficient than the other methods we have reviewed in earlier sections but is currently limited to single-qubit gates. Indeed, we show below that DB requires only four sets of state-fidelity measurements to accurately estimate all incoherent and coherent errors relevant to single-qubit gate operations. Additionally, DB can also be used to characterize the bath temperature.
Moreover, we show that DB also enables exploration of how coherent and incoherent errors interact to cause biased noise, whose impact on fault tolerance has been thoroughly studied~\cite{Aliferis:2008aa,XZZX2021}. 

The gate sequences used in DB resemble the simplest dynamical decoupling (DD) pulse sequences introduced decades ago in nuclear magnetic resonance \cite{Hahn:50, CP1954, CPMG1958}. But, while DD is nowadays typically used to \emph{suppress} incoherent and coherent noise during \emph{idling}---whether between gates \cite{Viola1998, Uhrig2007, Suter:2016aa, jurcevicDemonstrationQuantumVolume2021, pokharel2022demonstration, baumer2024quantum} or as a result of crosstalk \cite{Tripathi2022, Zhou2023, baumer2023efficient, Shirizly:2024aa, NIU2024, evert2024syncopated, brown2024efficient}, the goal of DB is instead to \emph{characterize} the noise arising during \emph{gate operation}.

\subsection{Qubit model}
We base our presentation on transmon qubits\cite{Koch2007}, which we model as driven two-level systems and consider in the drive frame under the rotating wave approximation~\cite{TripathiTransmon2024}. However, the DB methodology is general and applies to any qubit modality.

Let $R_\alpha(\theta)\equiv \exp[-i (\theta/2) \sigma_\alpha]$, where $\{\sigma_\alpha\}$ is the set of Pauli matrices. The Hamiltonian that generates the single-qubit gates $R_x(\theta)$ and $R_y(\theta)$, including both phase and rotation errors, is given by:
\begin{equation}
    H_\alpha = \left(\eps +  \epserr\right)\frac{\sigma_\alpha}{2} + \delerr  \frac{\sigma_z}{2} \ , \quad \alpha\in\{x,y\} .
    \label{eq:Hsys}
\end{equation}
Ideally, both the amplitude error $\epserr$ and the detuning error $\delerr$ vanish. In reality, both are present and give rise to rotation and phase errors
\begin{equation}
\dth\equiv \epserr  t_g\ , \quad  \dphi \equiv \frac{\delerr}{\bar{\eps}} = \frac{1}{\theta}\delerr t_g
\label{eq:dth-dphi}
\end{equation}
respectively, with $t_g$ denoting the gate duration, $\theta \equiv  \int_0^{t_g}\varepsilon(t)dt$, and $\bar{\varepsilon} = \theta/t_g$ the 
average pulse amplitude.
We expect to see a pulse-shape dependence in the phase error $\delta\phi$, as the effective detuning error $\delerr$ may depend on instantaneous drive amplitude $\varepsilon(t)$.
We assume square-shaped pulses for our analytical theory and cosine-shaped pulses for our numerical calculations using the Lindblad master equation (cosine pulses are used in experimental measurements).

\cref{eq:Hsys} is an effective model, where $\delerr$ is an error resulting from both detuning---a discrepancy between the drive and qubit frequencies---and the ac-Stark shift induced by driving other transitions (such as the higher levels of the transmon) off-resonantly during the gate \cite{Martinis2014}. 
The two-level model is justifiable, given that leakage to higher levels of the transmon during our gates is very small
(see \cref{sec:leakage}). 

\subsection{Four key parameters}
Recall from \cref{sec:QDM} that any completely positive trace-preserving (CPTP) map acting on a $D$-dimensional Hilbert space can be characterized using no more than $D^4-D^2$ real parameters, i.e., $12$ for a single qubit. While RB reduces this to a single effective parameter, 
DB makes a less drastic reduction, to four key parameters: the relaxation time $T_1$, the coherence time $T_2$ (which contains information on pure dephasing $T_\phi$), the rotation error $\dth$, and the phase error $\dphi$. Assuming a Markovian noise model, we show below that these parameters can be straightforwardly extracted and, at the same time, provide just the right amount of information about the errors affecting single-qubit gates to enable such gates to be accurately calibrated.  

We denote $X \equiv R_x(\pi)$, $\bX \equiv R_x(-\pi)$, and similarly for $Y$ and $\bY $. To define DB, we need gate-pair sequences of the form $P_1P_2$. In particular, we consider $P_1P_2\in \{XX, X\bX , YY, Y\bY , \bY Y\}$. Note that in the absence of any errors, i.e., when $\epserr = \delerr  = 0$ and without coupling to a bath, $P_1P_2=\pm I$ for this set (the global phase will not matter). In reality,  $P_1P_2\not\propto I$ and we consider the corresponding fidelity $\tilde{F}(n) =|\bra{\psi} (P_1 P_2)^n \ket{\psi}|^2$ where $\ket{\psi}$ is the initial state and $n\in \mathbb{Z}^{+}$ is the number of repetitions of the $P_1P_2$ sequence; hence the fidelity is defined stroboscopically. We use a tilde to denote the experimentally measured fidelity.

The core DB protocol consists of a set of four ``learning'' experiments designed to extract $\{T_1,T_2,\dth,\dphi\}$, followed by tests involving new experimental data. In each case, the notation $\{P_1P_2;\ket{\psi}\}$ means a circuit comprising preparation of the state $\ket{\psi}$ from the $\ket{0}$ state, a varying number of repetitions of the $P_1P_2$ sequence, unpreparation of $\ket{\psi}$, and finally a measurement in the $\sigma^z$ eigenbasis. Finally, the empirical fidelity $\tilde{F}$ is found for each experiment as the ratio of $\ket{0}$ outcomes to the total number of experimental shots ($800$ in our case). We find that in all these experiments---learning as well as testing---the fidelity can be accurately described using a simple analytical expression:
\begin{equation}
    F(t_n) = \frac12(1+a) +  \frac12(1-a)e^{-t_n/T_D}\cos(2\omega t_n) ,
    \label{eq:fid}
\end{equation}
where $t_n=2n t_g$ represents the total evolution time. The parameter $T_D$ denotes a characteristic timescale associated with incoherent noise processes, while $\omega$ is a frequency associated with coherent errors. The parameter $a$ quantifies the deviation from the infinite temperature limit, where 
$F(\infty)= \frac12$.
\cref{eq:fid} effectively captures the interplay between coherent and incoherent noise sources. While it is phenomenological, it is also derivable from a Markovian master equation model subject to certain approximations, as shown in
\cref{sec:fid_derivation}. 

Finally, we note that \cref{eq:fid} can be further generalized if we replace $\frac12(1-a)\mapsto A$ and $\frac12(1+a)\mapsto B$, which adds an extra parameter.  If we set $\omega=0$ (no coherent errors), this reduces to an RB fidelity-like expression $Ap^n+B$, where $p=e^{-2t_g/T_D}$ [recall \cref{eq:RB_decay}]. RB's ability to quantify state preparation and measurement (SPAM) errors as $1-(A+B)$ is thus captured by this generalization of DB; however, we do not consider this here, as our focus is the determination of $\{T_1,T_2,\dth,\dphi\}$. Finally, note that coherent errors introduced by leakage do not fit the form of \cref{eq:fid}; we discuss this further in \cref{sec:leakage}.

\subsection{Protocol}%
We now detail the DB protocol, which comprises the following experiments or steps:
\begin{enumerate}[leftmargin=*, itemsep=0pt, topsep=0pt, parsep=0pt]
    \item $\{\text{free};\ket{1}\}$: This is a standard $T_1$ measurement, where we do not apply any pulses (free evolution). \cref{eq:fid} reduces to $\frac12(1+a)+\frac12(1-a) e^{-t/T_1}$ in this case which we fit to determine $T_1$. This yields the relaxation component of the Lindbladian.
    
    \item $\{XX;\ket{+}\}$: Fitting this experiment with \cref{eq:fid} yields $T_D = T_2$, the coherence time from which we deduce the pure dephasing time $T_\phi =  2T_1 T_2/(2 T_1 - T_2)$. This is analogous to an echo-Ramsey measurement \cite{oliver_rev} and yields the dephasing component of the Lindbladian. This experiment is also sensitive to leakage (discussed below); here we assume leakage is small.

    \item $\{YY;\ket{+}\}$: This yields fidelity oscillations which we fit to find the rotation error $\dth$ where $\omega\approx \dth/(2t_g)$ in \cref{eq:fid}. 
    
    \item $\{X\bX ;\ket{+}\}$: Similarly, this yields fidelity oscillations which we fit to find the phase error $\dphi$ where $\omega = \dphi / t_g$ in \cref{eq:fid}. The decay times $T_D$ obtained here and in step 3 are a combination of the incoherent and coherent noise and hence different from $T_2$ obtained in step $2$. We now have all four parameters, which we use to numerically calculate predictions for the test experiments.
    
    \item Test 1: $\{Y\bY ;\ket{+}\}$ and $\{\bY Y;\ket{+}\}$: This highlights the interplay of asymmetric relaxation noise and gate operations and can also be used to characterize the qubit temperature
(see \cref{sec:finite_temp}). 
    
    \item Test 2: Various other experiments. Here, we report tests using the UR6 dynamical decoupling sequence \cite{Genov2017}.  
\end{enumerate}

Next, we demonstrate that this protocol enables a detailed assessment of both coherent and incoherent errors affecting single-qubit gates. 

\begin{figure}[t]
\centering
\includegraphics[width=0.47\textwidth]{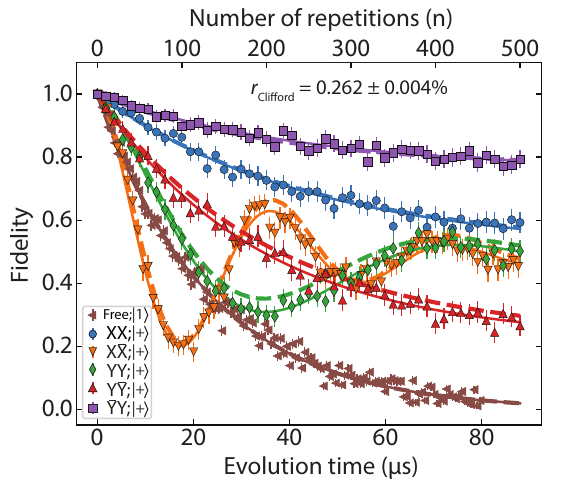}
\caption{Empirical ($\tilde{F}$, symbols), analytical fit ($F$, solid curves), and Lindblad-based ($\bar{F}$, dashed curves) fidelities for the initial states $\ket{1}$ under free evolution and $\ket{+}$ under various gate sequences $P_1P_2$. 
The analytical fit yields $T_1=23.36 \pm 0.40~\mu$s, $T_2 = 44.13\pm 2.49~\mu $s, $\dth =0.398\pm  0.004^\circ$, and $\dphi = 0.426\pm 0.004^\circ$. With the exception of the $X\bX$ (orange) and $YY$ (green) cases, the Lindblad-based results are indistinguishable from the analytical fits.}
\label{fig:dd_expt}
\end{figure}

\subsection{Experimental results}%
We conduct all our experiments using a superconducting transmon qubit dispersively coupled to its readout resonator 
(see \cref{app:device} for more details). 
We implement $R_y(\theta)$ gates by adjusting the phase of the microwave pulses used for $R_x(\theta)$ gates, thus eliminating the need for additional calibrations. 
All single-qubit gates are calibrated using the Derivative Removal of Adiabatic Gates (DRAG) technique \cite{Motzoi2009} to minimize leakage and phase errors induced by higher levels of the transmon. To emulate drifts in experimental parameters, we first perform measurements on an outdated set of gate calibration parameters. The results are shown in \cref{fig:dd_expt}.

We observe that $\tilde{F}_{YY}$ oscillates, indicating rotation errors $\dth$, as explained below. Similarly, $\tilde{F}_{X\bX}$ oscillates due to phase errors $\dphi$. The $XX$ sequence on $\ket{+}$ is insensitive to coherent errors and, instead, $\tilde{F}_{XX}$ exhibits an exponential decay characterizing the $T_2$ time during time-dependent gate operations. The $T_D$ parameter from the $XX$ sequence is chosen as the effective $T_2$ since, subject to continuous time evolution under the noiseless version of this sequence, starting from $\ket{+}$ the state always remains $\ket{+}$.
Also shown in \cref{fig:dd_expt} are results from the $\bY Y$ and $Y\bY $ sequences. We discuss the reason for the pronounced difference they exhibit below. 

We use \cref{eq:fid} to fit (solid curves) each of the four experimental fidelities (symbols) in \cref{fig:dd_expt} prescribed by steps 1-4 of the DB protocol and find close agreement. 

\begin{figure}[t]
		\centering
		\includegraphics[width=\columnwidth]{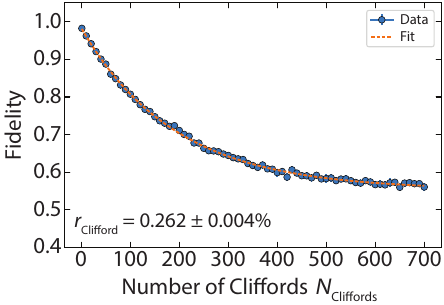}
		\caption{Randomized benchmarking measurement for the same gate calibration parameters used in \cref{fig:dd_expt}. The error bars represent the deviation of the averaged benchmarking results across 30 distinct random sequences with a maximum number of Cliffords ($N_\mathrm{Cliffords}$) of 700, which are smaller than the marker sizes.}
		\label{fig1:RB}
\end{figure}

We contrast the resulting $\{T_1, T_2, \dth, \dphi\}$ values with the single metric $r_C$ derived from the RB protocol [\cref{eq:rb_infid}]. \cref{fig1:RB} shows the average RB result for 30 distinct experiments with a total length of 700 Clifford gates for the same gate parameters shown in \cref{fig:dd_expt}. On average, our RB Clifford gate set consists of approximately $80\%$ $\pi/2$ and $20\%$ $\pi$ gates, with $\pi/2$ pulses expected to exhibit half the rotation errors and one-fourth of the phase errors. The RB curve can be fitted to a function of the form $A p^m + B$ (\cref{fig1:RB}). This process yields an average gate infidelity of $r_C = (1-p)/2  =0.262\pm 0.004 \%$ for the single-qubit Clifford gate set,
a far less detailed characterization that misses the coherent oscillations and the distinction between $T_1$ and $T_2$. 
This also shows that RB alone is not sufficient to achieve accurate gate calibration. Note that this method is limited by the relaxation time of the qubit ($T_1$) as well as the gate duration ($t_g$) with the average single-qubit gate fidelity scaling as $\sim t_g/T_1$ \cite{RB_2009} and in our case, the gate fidelity is significantly affected by the relatively short $T_1 \approx 20 \ \mu$s of the device.

As we show in the next several sections, all observed features can be understood using either a simple, analytically solvable closed-system model or a numerical model based on a Lindblad master equation with time-dependent gates. We first provide the analytical fidelities in the closed system case, which exhibit oscillations for certain sequences. 

\subsection{Fidelity with coherent errors}
Assuming the system is closed, \cref{eq:Hsys} can be used to analytically compute the fidelities of the gate sequences comprising 
our DB protocol. The closed system fidelity $F^{\text{clsd}}_{P_1P_2}(n)$ then reflects only the coherent errors contained within $P_1$ and $P_2$. For $YY$ and $XX$ and the initial state $\ket{+}$, we find:
\begin{subequations}
\begin{align}
\label{eq:Fyy} F^{\text{clsd}}_{YY}(n) &= \cos^2(n\therr) \\
\label{eq:Fxx} F^{\text{clsd}}_{XX}(n) &= 1 - \left( \frac{\pi\dphi \sin(n\therr)}{\therr}\right)^2,
\end{align}
\label{eq:4}
\end{subequations}
where $\therr \equiv \sqrt{ (\eps t_g+ \dth)^2+(\pi\dphi)^2}$.

Expanding to the leading order in $\dphi$ and $\dth$, with $\eps t_g= \pi$, $F^{\text{clsd}}_{YY}(n)$ simplifies to $ \cos^2(n \dth)$. Thus, \emph{the fidelity under the $YY$ sequence is dominated by rotation errors $\dth$}. 

Regarding $F^{\text{clsd}}_{XX}(n)$, in the small error limit, the second term in \cref{eq:Fxx} is suppressed, and $F^{\text{clsd}}_{XX}(n)\approx 1$. Consequently, there is no dependence on either the rotation or the phase error. Finally:
\begin{equation}
\label{eq:fid-XXb}
\begin{split}
    F^{\text{clsd}}_{X\bX }(n) &= 
\cos^2\left(n\varphi_{\mathrm{err}}\right)\\
\varphi_{\mathrm{err}}&\equiv \tan^{-1} \left(
	2\pi \frac{\dphi}{\therr}  \sin\left(\frac{\therr}{2}\right) \frac{\sqrt{1-\Lambda/2}}{1-\Lambda}
\right)\\
\Lambda &\equiv \left( \pi \frac{\dphi}{\therr} \right)^2 (1-\cos{\therr})
\end{split}
\end{equation}
Expanding to leading order again, with $\eps t_g=\pi$, we have $F^{\text{clsd}}_{X\bX }(n) \approx \cos^2\left[2n\dphi\left(1-\frac{1}{\pi}\dth\right)\right]$. Thus, \emph{the fidelity under the $X\bX$ sequence is dominated by phase errors $\dphi$}. 

These results justify our assignment of the oscillations observed in the $YY$ and $X\bX $ experiments shown in \cref{fig:dd_expt} to rotation and phase errors, respectively.

\subsection{Open system model}
The cosine dependence in \cref{eq:fid} is well motivated by the closed system (coherent error) fidelities given above, corresponding to $T_D=\infty$ and $a=0$. The choice of the two incoherent parameters in the DB protocol, specifically $T_1$ (associated with free evolution) and $T_2$ (associated with the $XX$ sequence), warrants further investigation through numerical simulation to better understand their contributions to the parameters $T_D$ and $a$ in \cref{eq:fid}. We use a simple Lindblad master equation $\partial_t\rho = \mc{L}(\rho)$ to capture the effects of non-unitary evolution, with a Lindbladian of the form 
\beq
\mathcal{L} = -i[H, \cdot] + \sum_{\alpha\in\{1,\phi\}} \gamma_\alpha (L_\alpha \cdot L_\alpha^{\dagger}-\frac{1}{2}\{L_\alpha^{\dagger} L_\alpha, \cdot\}).
\eeq 
The Lindblad operators are $L_1 = \sigma^{-} = \ketb{0}{1}$ and $L_{\phi} = \sigma_z/\sqrt{2}$ with rates $\gamma_1 = 1/T_1$ (relaxation) and $\gamma_{\phi} = 1/T_\phi =  (2 T_1 - T_2)/(2T_1 T_2)$ (dephasing). 

Using the fitted parameters $\{T_1, T_2, \dth, \dphi\}$ from Steps 1-4 of the DB protocol, we numerically solve the Lindblad equation and find good agreement with the experimental results, as shown by the dashed lines in \cref{fig:dd_expt}. Moreover, knowledge of these parameters allows us to proceed to step 5 of the protocol and predict, as a test, the outcomes of the $Y\bY$ and $\bY Y$ experiments. As shown in \cref{fig:dd_expt}, there is again good agreement between the simulations and the experimental results. This validates the choice of $\{T_1, T_2, \dth, \dphi\}$ as an appropriate set of parameters. In contrast, $r_C$ is clearly insufficient for predicting the outcomes of the $Y\bY$ and $\bY Y$ experiments.

\subsection{Sensitivity to coherent errors}

\begin{figure}[t]
\centering
\includegraphics[width=0.48\textwidth]{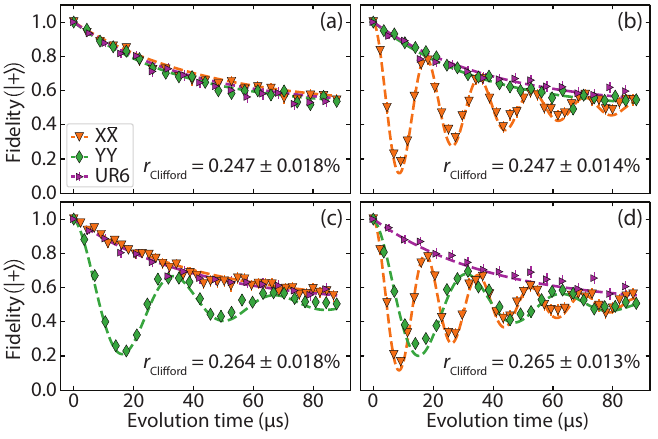}
\caption{Experimental fidelities $\tilde{F}^{\ket{+}}$ showing the very different sensitivity of DB and RB to coherent errors. (a) $\dth = 0^\circ$, $\dphi = 0^\circ$, (b) $\dth = 0^\circ$, $\dphi = 0.893\pm 0.002^\circ$, (c) $\dth = 0.932\pm0.007^\circ$, $\dphi=0^\circ$, (d)~$\dth = 0.995\pm 0.009^\circ$, $\dphi = 0.90\pm 0.002^\circ$. These angles are obtained from fits using \cref{eq:fid}. Dashed curves are the numerically computed fidelities $\bar{F}^{\ket{+}}$ from the open system Lindblad model using the DB parameters above along with the corresponding $T_1$ and $T_2$. The UR$6$ sequence suppresses even large coherent errors.} 
\label{fig:dd_errors}
\end{figure}

Having confirmed that DB effectively identifies various errors, we next assess the sensitivity of both DB and RB to coherent errors. Initially, we calibrate the gates to minimize both $\dphi$ and $\dth$ 
(see \cref{app:calib}), 
and immediately follow this calibration by a series of DD experiments, as depicted in \cref{fig:dd_errors}(a). These experiments focus on the $X\bX $ and $YY$ sequences, which are sensitive to phase and rotation errors. As a second test (Step 6 in the DB protocol), we include a Universally Robust (UR$n$) DD sequence, specifically designed to suppress coherent errors~\cite{Genov2017}. 
We then deliberately introduce coherent errors with ($\dth$, $\dphi$) values  given in \cref{fig:dd_errors}. The phase error $\dphi$, is introduced by mistuning the DRAG weighting parameter $\alpha$, which effectively varies the detuning error term $\delerr$ in our model.  The slight increase in rotation error $\dth$ from \cref{fig:dd_errors}(c) (in the absence of $\dphi$) to (d) (in the presence of $\dphi$) can be attributed to the second order dependence of $\therr$ on $\dphi$, given by $\frac{1}{2}(\pi\dphi)^2/(\pi+\dth)$. 

In each of cases (a)-(d) we also conduct RB experiments with the same values of $\delerr$ and $\dth$. More precisely, 
when we add coherent errors to the RB gates, they are added to both $R(\pi/2)$ and $R(\pi)$ in such a way that for $R(\pi/2)$ the rotation errors are $1/2$ that of $R(\pi)$ and the phase errors for $R(\pi/2)$ are $1/4$ of that of $R(\pi)$.

As anticipated, immediately following gate calibration using the parameters extracted from the DB protocol, neither $X\bX $ nor $YY$ causes oscillations, as shown in \cref{fig:dd_errors}(a), and the coherent error is negligible. However, after deliberately introducing phase errors, significant oscillations appear in $\tilde{F}_{X\bX}$, as depicted in \cref{fig:dd_errors}(b), while $\tilde{F}_{YY}$ simply decays exponentially without exhibiting any oscillations. In contrast, introducing a rotation error in \cref{fig:dd_errors}(c) reveals clear oscillations in $\tilde{F}_{YY}$ but none in $\tilde{F}_{X\bX}$. When both errors are introduced, \cref{fig:dd_errors}(d) displays clear oscillations for both sequences. 

The UR6 sequence never exhibits oscillations, confirming that coherent errors are their source. The Lindblad model with parameters obtained via the DB protocol (dashed curves) is again in good agreement with the experimental fidelities, further confirming the utility of DB.
In contrast, the error rate derived from RB, $r_C$, is unchanged when a phase error ($\dphi \approx 0.90^\circ$) is introduced and increases only slightly ($\approx 0.02\%$) with a rotation error ($\dth \approx 0.93^\circ$). We see that small coherent errors can cause large, rapid oscillations in sequence fidelity even when the RB fidelity is almost unchanged. This illustrates RB's limitations in detecting specific types of coherent errors. 

Finally, note that in the presence of only detuning error $\delerr$ in \cref{eq:Hsys}, the Rabi rate increases from $\eps$ to $\sqrt{\eps^2+\delerr ^2}$. Thus, for a gate time $t_g$, there would be an additional induced rotation error $\dth_{\rm ind} = (\sqrt{\eps^2+\delerr ^2}-\eps)t_g \simeq \delerr ^2t_g/(2\eps)$ (assuming a square pulse), even without a drive amplitude error. Using $\delerr = \bar{\eps} \dphi$ and $\bar{\eps} = \theta/t_g$, this gives
\begin{equation} 
\label{eq:inducedRotError}
    \dth_{\rm ind} \simeq \frac{1}{2}\phi\dphi^2.
\end{equation}
From the data in \cref{fig:dd_errors}, we have at most $\dphi = 0.90 ^\circ$ with a nominal rotation angle of $\theta = \pi$, giving $\dth_{\rm ind} \approx 0.02^\circ$ which is negligible. We therefore ignore the induced rotation error in our analysis.

\subsection{Interplay of $T_1$ asymmetry and gates} \label{sec:T1asymm}%
An intriguing observation arising from \cref{fig:dd_expt} is the strong asymmetry in the decay of the $Y\bY $ and $\bY Y$ fidelity curves. Transmon qubits operate at frequencies in the $3-6$ GHz range and temperatures around $10$ mK ($\sim 0.2$ GHz).
In this regime, the thermal energy is negligible, and the ground state $\ket{0}$ is approximately the thermal state of the transmon qubit, i.e., excitation from $\ket{0}$ to $\ket{1}$ is negligible and relaxation dominates. Tracing the Bloch sphere trajectories corresponding to different sequences reveals that $Y\bY $ confines the state to the southern (excited state) hemisphere, while $\bY Y$ limits it to the northern (ground state) hemisphere. Therefore, the former is more susceptible to relaxation ($T_1$) errors, which explains the faster decay and lower saturation fidelity of $Y\bY $ compared to $\bY Y$. This asymmetry is also well captured by the time-dependent Lindblad model, as shown by the dashed curves in \cref{fig:dd_expt}.

\begin{figure}[t]
		\centering
		\includegraphics[width=0.48\textwidth]{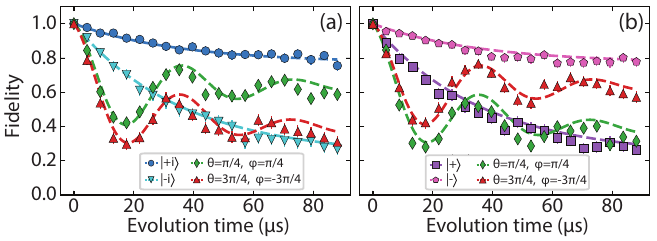}
		\caption{Asymmetry in the decay pattern of DB sequences due to the interplay of gates and $T_1$. Fidelity decay for (a) $X\bX$ and (b) $Y\bY$ applied to two pairs of orthogonal initial states. Dashed curves represent the Lindblad master equation simulation results with parameters obtained by fitting \cref{eq:fid}.}
		\label{fig:dd_t1}
\end{figure}

To demonstrate this effect, we performed additional experiments in the same outdated gate calibration cycle used in \cref{fig:dd_expt}. We applied the $X\bX $ sequence to $\ket{\pm i}$ (eigenstates of $\sigma_y$), as well as to $\ket{\psi_{\frac{\pi}{4},\frac{3\pi}{4}}}$ and $\ket{\psi_{\frac{3\pi}{4},-\frac{3\pi}{4}}}$, where $\ket{\psi_{\theta, \phi}} = \cos(\theta/2)\ket{0} + e^{i\phi}\sin(\theta/2)\ket{1}$. We observe a clear asymmetry between these pairs of orthogonal states under $X\bX $, as shown in \cref{fig:dd_t1}(a). We further confirm this asymmetry using $Y\bY $ applied to $\ket{\pm}$ (eigenstates of $\sigma_x$) and $\{\ket{\psi_{\frac{\pi}{4},\frac{3\pi}{4}}},\ket{\psi_{\frac{3\pi}{4},-\frac{3\pi}{4}}}\}$, finding the same asymmetric decay patterns both experimentally and numerically in \cref{fig:dd_t1}(b). These results illustrate the dependence of relaxation during time-dependent gates on the initial state and its trajectory under different pulse sequences. Numerical modeling of this asymmetry enables qubit temperature calibration.
(see \cref{sec:finite_temp}).

\subsection{Derivations: Fidelities of the $YY$ and $X\bX$ sequences}
\label{sec:fid_derivation}

In this section, we present a detailed analytical derivation of the fidelity subject to Markovian decoherence and coherent errors. We aim to derive our phenomenological fidelity expression, \cref{eq:fid}, as well the specific sequence dependence of the decay constant $T_D$, the coherent oscillation frequency $\omega$, and the SPAM parameters $A$ and $B$.

We assume a Markovian environment which we model via the Lindblad master equation with time-dependent driving: 
\begin{equation}
\begin{split}
\dot{\rho} &=-i\left[H(t), \rho\right]+\mathcal{L}_D(\rho)\\
 \mathcal{L}_D(\rho) &= \sum_\alpha \gamma_\alpha(L_\alpha \rho L_\alpha^{\dagger}-\frac{1}{2}\{L_\alpha^{\dagger} L_\alpha, \rho\}) . 
\end{split}
\end{equation}
We recognize that this is a phenomenological model and that rigorously derived master equations for time-dependent driving have a different structure, where the Lindblad operators $L_\alpha$ become time-dependent~\cite{ABLZ:12-SI,Dann:2018aa,Venuti:2018aa,Mozgunov:2019aa,nathan2020universal,Davidovic2020completelypositive,Davidovic:2022aa}. However, this simplified model is analytically solvable and suffices for our purposes.

\subsubsection{Background}
For analysis purposes, we work in the Bloch vector picture, which substitutes $\rho = \frac12 (I + v_x \sigma_x + v_y \sigma_y + v_z \sigma_z)$, $H = \frac12 (h_x \sigma_x + h_y \sigma_y + h_z \sigma_z)$, where $\bv = (v_x,v_y,v_z)\in \mathbb{R}^3, \bh = (h_x,h_y,h_z) \in \mathbb{R}^3$.
The dissipator $\mathcal{L}_D$ is replaced by $(R, \bc)$ with $R_{ij} \equiv \frac12\Tr (\sigma_i \mathcal{L}_D(\sigma_j))$ and $c_i \equiv \frac12\Tr (\sigma_i \mathcal{L}_D(I))$.
The Lindblad master equation then is $\dot{\bv} =  \bh \times \bv + R\bv + \bc$ (see, e.g., Ref.~\citenum{ODE2QME} for derivations of all these results).
The corresponding fidelity with the $\ket{+}$ state is given by 
\begin{equation}
\label{eq:F-vx}
F(t) = \frac12 (1+v_x(t)) .
\end{equation}

We model our system as a qubit with an energy gap $\omega_{01}$ coupled to a thermal environment at inverse temperature $\beta$ along with dephasing noise.
The combined noise model translates to $R = \text{diag}(-\gamma_{2},-\gamma_{2},-\gamma_1)$ and $\mathbf{c} = (0,0,\eta\,\gamma_1)$ where 
\begin{equation}
\gamma_{2} = \frac12 \gamma_1 + \gamma_\phi ,
\end{equation} 
and 
\begin{equation}
\label{eq:eta}
\eta = \frac{1-\exp(-\beta \omega_{01})}{1+\exp(-\beta \omega_{01})} \in [0,1] .
\end{equation}

\subsubsection{$YY$ sequence} 
\label{sec:YY-fid-proof}

For a square $YY$ pulse, $ \bh(t) = (0, \epstot, \delerr)$ where $\epstot = \eps+\epserr$. Then the Bloch vector equation becomes $\dot{\bv} = G\bv+\bc$, where
\begin{equation}
\label{eq:G}
G = \begin{pmatrix}
-\gamma_2 & -\delerr & \epstot \\
\delerr & -\gamma_2 & 0 \\
-\epstot & 0 & -\gamma_1
\end{pmatrix}.
\end{equation}
We first solve the case where $\delerr = 0$ ($\dphi=0$). Define
\begin{equation}
\label{eq:deltaG}
\delta G = \begin{pmatrix}
0 & -\delerr & 0 \\
\delerr & 0 & 0 \\
0 & 0 & 0
\end{pmatrix}.
\end{equation}

\paragraph{Exact solution for $\delerr = 0$}

The Bloch vector equation for $\delerr = 0$ becomes
\begin{equation}
\dot{\bv} = (G-\delta G)\bv+\bc.
\label{eq:simplified-YY}
\end{equation}
In this equation, $v_y$ is decoupled from the other two components and to solve for $v_x(t)$ and $v_z(t)$, requires exponentiation of a $2\times2$ matrix.
Given that $\bv(0) = (1,0,0)$, the relevant solutions for the Bloch vector components take the following form:
\begin{equation}
\begin{split}
v_y(t) &= 0 \\
\label{eq:simplified-YY-x}
v_x(t) &= v_\infty + (1- v_{\infty}) e^{-\gamma_*t} \left[ 
	 \cos{(\omega_{*} t)} +\vphantom{\frac{\gamma}{\omega}} \right. \\ 
	 & \left.   \left( \frac{(\gamma_1-\gamma_2) - (\gamma_1+\gamma_2)(1-v_\infty)}{2\omega_{*}(1-v_\infty)} \right) \sin{(\omega_{*} t)} 
     \right] \notag
\end{split}
\end{equation}
where 
\begin{equation}
\label{eq:11}
\begin{split}
v_\infty &= \eta \frac{\epstot \gamma_1}{\epstot^2+\gamma_1\gamma_2}\\
\gamma_{*} &= \frac{1}{2}(\gamma_1+\gamma_2)\\
\omega_{*} &= \sqrt{\epstot^2 - \frac14 (\gamma_1-\gamma_2)^2}\ .
\end{split}
\end{equation}
Along with \cref{eq:F-vx}, \cref{eq:simplified-YY-x} is close to, but not quite \cref{eq:fid}.

\paragraph{Perturbation theory for $\delerr = 0$}

In the regime where $\eps \gg \epserr, \gamma_1, \gamma_2$, the quantities in \cref{eq:11} simplify to 
\begin{equation}
v_\infty \approx \eta\frac{\gamma_1}{\eps}\ ,\quad \omega_{*} \approx \epstot = \eps + \epserr ,
\end{equation} 
to first order.
Moreover, since we are interested in the fidelities after each DD sequence, i.e., at $t=2nt_g$ ($n\in \mathbb{N})$, $\sin{(2\omega_{*}nt_g)} \approx \sin{(2n \dth)}$ and $\cos{(2\omega_{*}nt_g)} \approx \cos{(2n \dth)}$.
The term corresponding to $\sin{(2\omega_{*}nt_g)}$ vanishes to first order initially because it is multiplied by another first-order term and later because it is exponentially suppressed.
This results in a simplified expression for $v_x(2nt_g)$:
\begin{equation}
v_x(2nt_g) = v_\infty + (1- v_{\infty}) e^{- 2nt_g \gamma_*} \cos{(2n \dth)}
\end{equation}
The fidelity of the $\ket{+}$ state as a function of $YY$ repetitions $n$,  $F_{\dphi = 0}(n)$, is therefore given by
\begin{equation}
\begin{split}
F_{\dphi = 0}(t_n) &= \frac{1}{2}(1+v_\infty) +\\
&\qquad\frac12 (1- v_{\infty}) e^{-\gamma_* t_n} \cos(2n \dth) .    
\end{split}
\label{eq:fid-YY-dphi-0}
\end{equation}
\cref{eq:fid-YY-dphi-0} is identical in form to \cref{eq:fid}, with $a=v_\infty \approx \gamma_1/\eps$, $T_D=1/\gamma_*$ and $2\omega t_g \approx \dth$. Note that the oscillation period of $2n\dth$ matches that of the fidelity in the closed system case as discussed below \cref{eq:Fxx}.

\paragraph{Perturbation theory for $\delerr \ne 0$}

If the solution to the Bloch vector equation given in \cref{eq:simplified-YY} is denoted by $\bv_0 (t)$, then in the case where $\delerr \neq 0$, let the deviation of the actual Bloch vector trajectory $\bv(t)$ from $\bv_0(t)$ be given by $\bx(t) \in \mathbb{R}^3$, i.e., $\bv(t) =\bv_0(t) +  \bx(t)$.
Substituting this form into the Bloch vector equation $\dot{\bv} = G\bv+\bc$ gives
\begin{equation}
\dot{\bx} = G\bx + \delta G  \bv_0 .
\end{equation}
The solution for $\bx(t)$ is then given by
\begin{equation}
\bx(t) = \int_0^t e^{G(t-s)} \delta G \bv_0 (s) \,ds .
\end{equation}
Taking the norm of both sides and moving the norm inside the integral, we obtain
\begin{equation}
\norm{\bx(t)} \leq \int_0^t \norm{e^{G(t-s)}}  \norm{\vphantom{e^G}\delta G\bv_0 (s)} \,ds.
\end{equation}
Using \cref{eq:deltaG}, $\norm{\delta G \bv_0 (s)}$ simplifies to $\abs{ \delerr v_{0,x}(s)}$ since $v_{0,y}(s)=0$.
From \cref{eq:simplified-YY-x}, $\norm{v_{0,x}(s)}$ in turn can be strictly upper bounded by $\abs{v_\infty} + ke^{-\gamma_* s}$ for some constant $k$.
A smaller value of $k$ will result in a tighter bound.
In the approximation regime discussed above, $k$ can conveniently be chosen as  $2$.

If $G = SDS^{-1}$ where $D$ is the diagonal matrix of $G$'s eigenvalues $\lambda_i$ and $S$ is the similarity matrix with the corresponding eigenvectors of $G$ as its columns, then
\begin{equation}
\norm{e^{G(t-s)}} \leq \norm{S\vphantom{S^-1}}\norm{e^{D(t-s)}}\norm{S^{-1}} \leq \kappa e^{-\lambda (t-s)}
\label{eq:exp-norm}
\end{equation} 
where $\kappa = \norm{S\vphantom{S^-1}} \norm{S^{-1}}$ and $\lambda = -\max_i \Re(\lambda_i)$~\cite{Perko2001}. This gives the inequality
\begin{equation}
\label{eq:xi-bound}
\norm{\bx(t)} < 
     \begin{cases}
      \kappa \abs{\delerr}\left(\frac{1-e^{-\lambda t}}{\lambda}\abs{v_\infty} + k\frac{e^{-\gamma_* t}-e^{-\lambda t}}{\lambda-\gamma_*}\right) \\
      \hfill\text{ if}\ \lambda \neq \gamma_* \\
      \kappa \abs{\delerr}\left(\frac{1-e^{-\lambda t}}{\lambda}\abs{v_\infty} + k t e^{-\lambda t}\right)  
      \\ \hfill \text{otherwise}
    \end{cases}
\end{equation}
Note that the first term is $O(\delerr v_\infty/\lambda)$.

From first-order eigenvalue perturbation theory, we have
\begin{equation}
	\delta{\lambda_i} = \frac{\mathbf{w}_i \cdot\delta{G}\bv_i}{\mathbf{w}_i\cdot \bv_i} ,
\end{equation}
where $\mathbf{w}_i$ and $\bv_i$ are the left and right eigenvectors of $G$ in \cref{eq:G} with eigenvalues $\{\lambda_i\} = \{\-\gamma_*\pm i \omega_*,-\gamma_2\}$. Therefore, using \cref{eq:deltaG}:
\begin{equation}
\begin{split}
\lambda_{1,2} &= (-\gamma_* \pm i\omega_*) + O(\delerr) \\
\lambda_3 &= -\gamma_2  + O(\delerr) .
\end{split}
\end{equation}
Under the additional assumption $\delerr/\gamma_{1,2}=O(1)$, the first summand in \cref{eq:xi-bound} simplifies to $O(v_\infty) = O(\gamma_1/\eps)$.
Combining this with the fact that the second summand decays exponentially with time, $\xi(t)$ is a small enough quantity such that $\bv(t)$ can be well approximated by $\bv_0(t)$.
Therefore, the resulting fidelity expression $F_{YY}(n)$ in a $\delerr \neq 0$ system is still well captured by \cref{eq:fid-YY-dphi-0}.
It matches \cref{eq:fid} with an error $\abs{F_{YY}(n) - F_{\dphi = 0}(n)} < \frac12\norm{\bx(2nt_g)}$, with the factor $1/2$ due to \cref{eq:F-vx}. In big-$O$ notation,
\begin{equation}
\abs{F_{YY}(n) - F_{\dphi = 0}(n)} = O(\gamma_1/\eps) = O(\gamma_1 t_g) .
\end{equation}

\subsubsection{$X\bX$ sequence}
\label{sec:XXb-fid-proof}

We solve for a simplified model in which $T_{\phi} = 2T_1$, that is, $\gamma_1 = \gamma_2$: all the components of the Bloch vector decay uniformly to the steady state $(0,0,\eta)$.
The resulting Bloch vector equations are:
\begin{equation}
\dot{\bv}(t) =
     \begin{cases}
      G_+ \bv(t) + \bc,\\ \qquad \text{if}\ (2k)t_g < t \leq (2k+1)t_g,\,k\in \mathbb{Z} \\
      G_- \bv(t) + \bc,\\ \qquad \text{if}\ (2k-1)t_g < t \leq (2k)t_g,\,k\in \mathbb{Z} \\
    \end{cases}
    \label{eq:xxb-Bloch}
\end{equation}
where $\bc = (0,0,\eta\,\gamma)$, $\gamma = \gamma_{1}= \gamma_{2}$, and the matrices $G_\pm$ are
\begin{equation}
G_{\pm} = \begin{pmatrix}
-\gamma & -\delerr & 0 \\
\delerr & -\gamma & \mp \epstot \\
0 & \pm\epstot & -\gamma
\end{pmatrix}.
\end{equation}
The solutions for the Bloch vector after each $X\bX$ sequence (which corresponds to a duration of $2t_g$) are then given recursively by
\begin{equation}
\bv(2(n+1) t_g) = A \bv(2nt_g) - \bz
\end{equation}
where $\bz \in \mathbb{R}^3$, and
\begin{equation}
    \begin{split}
        A &= e^{G_- t_g} e^{G_+ t_g}\label{eq:29}\\
        \bz &= (I-e^{G_- t_g})(G_{-})^{-1}\bc + \\
        &\qquad e^{G_- t_g}(I-e^{G_+ t_g})(G_{+})^{-1}\bc.
    \end{split}
\end{equation}
Unraveling the recursion yields
\begin{equation}
\bv(2nt_g) = A^n \bv(0) - \sum_{k=0}^{n-1} A^k\bz . 
\end{equation}
We are interested in calculating only $v_x(2nt_g)$:
\begin{equation}
v_x(2nt_g) = (1,0,0)\cdot A^n \bv(0) - \sum_{k=0}^{n-1} (1,0,0)\cdot A^k\bz
\label{eq:xx-vx-complex}
\end{equation}
The required elements of the matrix $A^n$ can be written as:
\begin{equation}
\begin{split}
(A^n)_{11} &= \frac12 (p^{2n}+q^{2n})\\
(A^n)_{12} &= \frac{-i\omega'(1+e^{i\omega't_g})}{2\chi} (p^{2n}-q^{2n})\\
(A^n)_{13} &= \frac{-\epstot(1-e^{i\omega't_g})}{2\chi} (p^{2n}-q^{2n})
\end{split}
\end{equation}
where
\begin{equation}
\begin{split}
 \omega' &\equiv\sqrt{\epstot^2+\delerr^2} \\
 \chi &\equiv \sqrt{4e^{i\omega't_g}\omega'^2+(1-e^{i\omega't_g})^2\delerr^2} \\
 p &\equiv \frac{e^{-\gamma t_g - i\omega't_g}}{2\omega'^2} \left( (1+e^{2i\omega't_g})\delerr^2 + 2e^{i\omega't_g}\epstot^2 + \right. \notag\\
 &\qquad\qquad \left.(1-e^{i\omega't_g})\delerr \chi \right)\\
 q &\equiv \frac{e^{-\gamma t_g - i\omega't_g}}{2\omega'^2} \left( (1+e^{2i\omega't_g})\delerr^2 + 2e^{i\omega't_g}\epstot^2 -\right. \notag \\
 &\qquad\qquad\left.(1-e^{i\omega't_g})\delerr \chi \right) .
\end{split}
\end{equation}
Substituting these into \cref{eq:xx-vx-complex} gives
\beq
\begin{split}
& v_x(2nt_g) = \frac12 (p^{2n}+q^{2n}) +
\\ & \frac{1-p^{2n}}{2(1-p)} \left( \zeta_x - \frac{i(1+e^{i\omega' t_g})\omega'}{\chi} \zeta_y \right.\notag
\\ &\qquad\qquad\qquad\qquad \left. - \frac{(1-e^{i\omega' t_g})\epstot}{\chi}\zeta_z \right) + \notag
\\  &\frac{1-q^{2n}}{2(1-q)} \left( \zeta_x + \frac{i(1+e^{i\omega' t_g})\omega'}{\chi} \zeta_y + \right.\notag 
\\ &\qquad\qquad\qquad\qquad \left. \frac{(1-e^{i\omega' t_g})\epstot}{\chi}\zeta_z \right) , \notag
\end{split}
\eeq
where $\bz$ carries the temperature dependence. Under first order approximations, $\chi \approx 2e^{i\omega't_g/2}\omega'$, $\omega' \approx \epstot$, and
\begin{equation}
\begin{split}
 p & \approx e^{-\gamma t_g-i\vartheta}
\\  q & \approx e^{-\gamma t_g+i\vartheta}
\\  \vartheta &\equiv \tan^{-1}{\left( 2\sin{\left( \frac{\omega' t_g}{2} \right)} \frac{\delerr}{\omega'} \right)} .
\end{split}
\end{equation}
The values of $\bz$'s components are lengthy expressions and are given in \cref{app:zeta} so as not to interrupt the flow. Under first-order approximations, $\zeta_x \approx 0$, $(1+e^{i\omega' t_g})\zeta_y/\chi  \approx 0$ and $\zeta_z \approx 0$. Substituting these values gives the following final expression:
\begin{equation}
\begin{split}
v_x(2n t_g) &\approx \frac12 (p^{2n}+q^{2n})\\
&= e^{-2n\gamma t_g} \cos{2n\vartheta} .
\end{split}
\end{equation}
The expression for the fidelity with the $\ket{+}$ state is, therefore 
\begin{equation}
F(t_n) = \frac12  + \frac12 e^{-\gamma t_n} \cos{(2n\vartheta)} . \notag
\end{equation}
This expression matches \cref{eq:fid} with $2\omega t_g =\vartheta$, $T_D=1/\gamma$ and $a=0$.
Moreover, the oscillation frequency matches the closed system fidelity oscillation frequency [\cref{eq:fid-XXb}] up to the first order (noting that $\therr = \omega' t_g$).

Solving the general case where $\gamma_1 \neq \gamma_2$ is difficult: we cannot apply perturbative techniques directly to bound the deviation of the Bloch vector trajectory in a $\{\gamma_1\neq\gamma_2$,  $\delerr\neq0\}$ environment (where $v_x$ exhibits oscillating exponential decay) from a Bloch vector trajectory in a $\{\gamma_1\neq\gamma_2$, $\delerr = 0\}$ environment (where $v_x$ exhibits exponential decay). We rely instead on numerical simulations of the Lindblad equation.

\subsection{Finite temperature}
\label{sec:finite_temp}
We discuss numerical, experimental, and analytical approaches for accounting for finite-temperature ($T>0$) effects.

\subsubsection{Numerical}

In our simulations, we take the zero temperature limit, which fits our experimental data well and is typically valid for transmons. However, it is straightforward to generalize to a non-zero temperature. For zero temperature, we use a Lindblad operator $L_1 = \sigma^- = \ketb{0}{1}$ with rate $\gamma_1 = 1 / T_1$. For a non-zero temperature $T$, we replace this with two operators $L_+ = \sigma^- = \ketb{0}{1}$ and $L_- = \sigma^+ = \ketb{1}{0}$, with rates $\gamma_-$ and $ \gamma_+$, respectively. The rates obey $\gamma_-+\gamma_+ = \gamma_1 = 1 / T_1$ and the detailed balance condition $\gamma_+ / \gamma_- = \mathrm{exp}(-\hbar \omega/k_B T)$~\cite{Breuer:book}, where $\omega$ is the qubit frequency. Varying $T$ in our numerical model thus allows us to fit dynamics at any temperature. In particular, relaxation-sensitive sequences, such as $Y\overline{Y}$ applied to $\ket{+}$, saturate to a fidelity that strongly depends on the device temperature.

\subsubsection{Experimental}

The effect of finite temperature can also be seen in our DB protocol. In general, state preparation and measurement (SPAM) errors will cause the initial state fidelity to be less than 1. We can calibrate these errors out by defining the last point in the $T_1$ sequence (i.e., the thermally equilibrated state) to be a fidelity of $0$, and the first point (i.e., the population-inverted thermal state) to be a fidelity of $1$. Such a definition accounts for all temperature effects in state preparation and all measurement errors, rescaling all fidelities by the same multiplicative factor, and thus leaving the fitting and analysis unaffected. If we then compare the saturation fidelities of the $XX$, $Y\bY $, and $\bY Y$ sequences to the midpoint of the $T_1$ fidelities, we see that they differ in a way that depends on the temperature, all saturating to a fidelity of $0.5$ in the limit of infinite temperature. We can thus use DB to extract the device temperature, although using the $XX$ sequence (with saturation close to the mixed state) will be less sensitive to small temperature variations than using $Y\bY$ and $\bY Y$, or typical Rabi-based methods \cite{geerlingsDemonstratingDrivenReset2013}. Nevertheless, our original 4-step protocol captures the effect of finite temperature. Note that this technique is insensitive to measurement error, which can be significant and would disrupt temperature measurements based on single-shot state-assignment readout.

\subsubsection{Analytical}

Given $a$ in \cref{eq:fid} and square pulses, in fact, the temperature can be computed without simulation of the Lindblad master equation. This can be shown in the following way: suppose that the Bloch vector equation is the same as \cref{eq:xxb-Bloch}, but with $G_+$ and $G_-$ replaced with $G_1$ and $G_2$, respectively. $G_1$ and $G_2$ correspond to the two pulses chosen for the DD sequence; e.g., in an $XX$ sequence, $G_1 = G_2 = G_+$. Note that $G_{1}$ and $G_2$ carry no temperature dependence. Then, similar to the analysis in \cref{sec:XXb-fid-proof}, the Bloch vector after $n$ pulse-pairs is given by
\begin{equation}
\bv(2nt_g) = B^n \bv(0) - \eta \sum_{k=0}^{n-1} B^k\bz' ,
\label{eq:bloch-temp}
\end{equation}
where similar to \cref{eq:29},
\begin{equation}
\begin{split}
B &= e^{G_2 t_g} e^{G_1 t_g} \\
\bz' &= (I-e^{G_2 t_g})(G_{2})^{-1}\bc \\
&\qquad\qquad + e^{G_2 t_g}(I-e^{G_1 t_g})(G_{1})^{-1}\bc, \notag
\end{split}
\end{equation}
and 
$\bc = (0,0, \gamma_1)$.
The fidelity is then
\begin{equation}
\begin{split}
F(n) &= \frac{1 + \bv(0)\cdot B^n \bv(0) - 
\eta \sum_{k=0}^{n-1} \bv(0)\cdot B^k \bz'}{2}  
\end{split}
\end{equation}
where again $\eta$ is the thermal factor defined in \cref{eq:eta}.
For $n\rightarrow\infty$, $\norm{B^n}\rightarrow 0$ from \cref{eq:exp-norm} and
\begin{equation}
F(\infty) = \frac12\left(1 - \eta \bv(0)\cdot (I-B)^{-1} \bz'\right) ,
\end{equation}
where we used $\sum_{k=0}^{\infty} B^k = (I-B)^{-1}$. Substituting $F(\infty) = (1+a)/2$ from \cref{eq:fid} gives
\begin{equation}
a = \eta \bv(0)\cdot (B-I)^{-1} \bz' .
\label{eq:temperature}
\end{equation}
The LHS of \cref{eq:temperature} is measured experimentally, while the inner product in the RHS is computed using the parameters $\{T_1, T_2, \epstot, \delerr\}$. 

Note that $a$ is maximized when $\eta=1$, i.e., at zero temperature [\cref{eq:eta}]. 
Accordingly, substituting $a_{T = 0} = (\bv(0)\cdot (B-I)^{-1} \bz')$ into \cref{eq:temperature} and rearranging \cref{eq:eta} gives the temperature as
\begin{equation}
k_B T = \omega_{01}\left(\ln{\left( \frac{a_{T = 0} + a}{a_{T = 0}-a}\right)}\right)^{-1} .
\end{equation}
As mentioned above, for transmons $\omega_{01}\in[3-6]\,$GHz and $T\sim 10$ mK, so $a \lesssim a_{T = 0}$. However, for devices where finite temperature effects are significant, $a < a_{T=0}$; moreover $a \approx 0$ for $T \rightarrow \infty$.

Since the experimental error in calculating $a$ is largely unaffected by the type of pulse sequence (i.e., by the value of $a$ itself), it is advisable to select pulse sequences that have a higher value of $|a_{T = 0}|$ to minimize the fractional uncertainty in temperature $T$. 
E.g., a $\overline{Y}Y$ sequence which gives a simulated value of $\abs{a_{\overline{Y}Y}}$ would be preferred over a $YY$ sequence which gives $\abs{a_{YY}}$ if $\abs{a_{\overline{Y}Y}} > \abs{a_{YY}}$.

\subsection{Leakage}
\label{sec:leakage}
\begin{figure}[t]
    \includegraphics[width=0.47\textwidth]{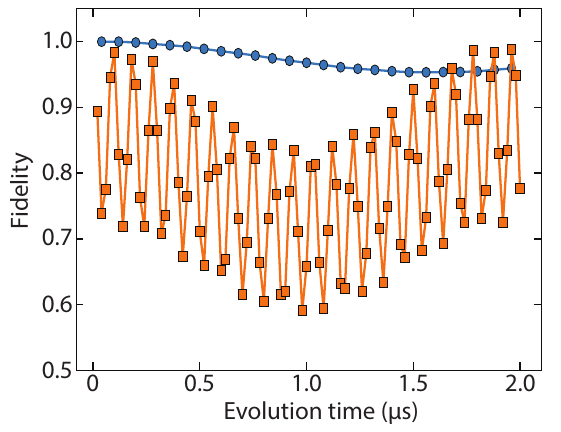}
    \caption{\label{fig:leakage} Simulated survival fidelity of the $\ket{+}$ state after repetitions of the $XX$ sequence, where each gate is a cosine pulse length of 20 ns (blue circles) or 10 ns (orange squares). A third state in the simulation represents leakage, with a detuning (anharmonicity) of -150 MHz relative to the qubit transition. Both the amplitude and frequency of oscillations increase as the gate duration is decreased and leakage becomes more significant.}
\end{figure}

Leakage to a state outside the qubit subspace is a coherent error as long as the leakage state's coherence is long-lived compared to the gate duration. The coherent rotation to the leakage state will lead to oscillations of the fidelity. These oscillations will, in principle, show up with any pulse sequence, but they are difficult to distinguish from the oscillations due to rotation errors in the $YY$ sequence or phase errors in the $X\bar{X}$ sequence. For this reason it is best to look at the $XX$ sequence, which has no oscillations in the absence of leakage. In this case the functional form \cref{eq:fid} no longer holds, as the amplitude of the oscillations and the amplitude of the decay are decoupled from each other. Because leakage is a dynamical effect that depends on the pulse shape, and because the leakage state may influence the readout signal, no simple analytical formula can capture the effect of leakage on fidelity. Qualitatively, however, it is easy to see that more leakage leads to larger and faster fidelity oscillations. For a simple example, see \cref{fig:leakage}. Here we have done a closed-system simulation of a qubit with a leakage state that is -150 MHz detuned from the qubit transition and has a drive matrix element $\sqrt{2}$ times larger---i.e., a qutrit similar to a transmon\cite{Goss:2022aa,tripathi2024quditdynamicaldecouplingsuperconducting}. The simulated pulses have cosine envelopes with durations of 10 or 20 ns (orange squares and blue circles, respectively). When prepared in $\ket{+}$ and driven with the $XX$ sequence, fidelity oscillations appear, with amplitude and frequency that increase as the gate time is decreased (gate bandwidth is increased).  Note that these are pulses without DRAG\cite{Motzoi2009} and so leakage can be quite significant. Adding DRAG with appropriate parameters can remove these oscillations. In our experiments the pulses were much slower relative to the anharmonicity and DRAG removed any residual leakage, so we ignore leakage in the experimental analysis. 

Because the $XX$ sequence on $\ket{+}$ is unaffected by rotation and phase errors, fidelity oscillations in $XX$ provide a direct diagnostic of the amount of leakage. This is particularly useful in the case where the leakage state is unknown and so measurement of the leaked population may not be possible. While no easy analytical solution presents itself, it should be possible to numerically model a gate sequence and fit the measured data to extract parameters of the leakage state.

\subsection{DB summary and future research directions}
DB seeks to strike a balance between the amount of information collected about the noise and efficiency of the process to obtain that information.
Moreover, DB highlights noise bias, which might benefit the development of compilation strategies~\cite{Krishnan-Vijayan:2024aa} provided circuits can be compiled so as to minimize relaxation by steering quantum trajectories away from excited states. The benefits could extend to resource estimation for fault-tolerant quantum algorithms \cite{beverland2022assessing, agrawal2024} and biased noise error correction \cite{XZZX2021}. 
The implementation of the DB protocol is straightforward and adaptable beyond superconducting qubits. We anticipate that DB will contribute to the benchmarking and enhancement of quantum gates across various platforms. A natural open problem is to extend DB to two-qubit gates and identify the corresponding minimal extended parameter set that describes incoherent and coherent errors.

\section{Conclusions and outlook} 
\label{sec:concl}
Gate benchmarking techniques are crucial for practical quantum computing, fulfilling two key roles: (1) offering realistic estimates of attainable performance with existing quantum hardware, and (2) supplying valuable metrics for assessing and monitoring advancements in various gate and circuit implementations. These methods differ widely in their assumptions about the computation and error models, their implementation's simplicity and cost, and the level of detail they provide about the noise and errors impacting the gates.

This review provides a comprehensive analysis of key benchmarking methodologies for assessing noise in quantum gates and circuits, emphasizing the trade-offs inherent in each strategy. Among the techniques explored are \textit{Randomized Benchmarking} (RB), well-known for its efficient and scalable computation of average gate fidelity; \textit{Quantum Process Tomography}, which, while entailing greater implementation costs, yields a complete noise characterization of gates but is sensitive to SPAM; \textit{Gate Set Tomography}, which has all the ingredients of quantum process tomography but includes state preparation and measurement errors explicitly in its model; \textit{Process Fidelity Estimation} and \textit{Direct Fidelity Estimation}, which deliver practical fidelity measures with reduced resource demands and are applicable to both circuits and individual gates; as well as \textit{Cross Entropy Benchmarking} and \textit{Random Circuit Sampling}, which are predominantly employed for evaluating complete quantum circuits, particularly in the context of demonstrating quantum supremacy. 
We also presented a new \emph{Deterministic Benchmarking} (DB) protocol and demonstrated it using a superconducting transmon qubit.
DB is deterministic and efficient: it employs a small and fixed set of simple pulse-pair sequences. The protocol quantifies both incoherent and coherent gate errors that elude RB and does so without averaging over random circuits, in contrast to all the other methods we surveyed, with the exception of QPT.

As quantum computing technology reaches maturity, benchmarking will become a valuable tool for fine-tuning these systems. Currently, however, it is essential for identifying noise and decoherence sources, as well as aiding in the development and optimization of quantum information devices until they achieve reliability and functionality. Ongoing enhancement and advancement of benchmarking methods are pivotal for raising the fidelity of quantum logic gates and circuits to meet the demands of scalable quantum computing architectures.

Ultimately, combining refined noise characterization and error mitigation holds the key to unlocking challenging quantum chemistry problems that extend beyond the scope of classical computation~\cite{Lee:2023aa}. In the near term, these improvements enable the design of noise-resilient ans{\"a}tze for VQE, thereby raising the ceiling on the system sizes that can be reliably tackled on noisy intermediate-scale quantum (NISQ) devices~\cite{Cao:2019aa}. Over the longer horizon, as fault-tolerance thresholds are approached and eventually surpassed, the careful measurement of device-specific noise will continue to inform quantum error correction procedures, bridging the gap between hardware capabilities and the exacting precision required for practical chemical applications. By continuing to improve noise characterization and error mitigation, we inch closer to a regime where quantum hardware can deliver transformative insights into the molecular processes and material properties that govern our physical world.

\textit{Acknowledgement}--- We thank Namit Anand, Nic Ezzell and Alexandru Paler for useful discussions. Devices were fabricated and provided by the Superconducting Qubits at Lincoln Laboratory (SQUILL) Foundry at MIT Lincoln Laboratory, with funding from the Laboratory for Physical Sciences (LPS) Qubit Collaboratory. This research was supported by the National Science
Foundation, the Quantum Leap Big Idea under Grant No.
OMA-1936388, the Army Research Office MURI Grant No. W911NF-22-S-0007, and
the Intelligence Advanced Research Projects Activity (IARPA) under Cooperative Agreement No. W911NF-23-2-0216. 

\appendix
\begin{center}
\Large \bf Appendix
\end{center}

This appendix describes technical details relevant to the Deterministic Benchmarking protocol.

\section{Device design and experimental setup}
\label{app:device}

The device used in the experiment consists of a simple grounded Xmon geometry \cite{Barends2013a} dispersively coupled to its co-planar waveguide (CPW) readout resonator. It is nominally identical to the device reported in Ref.~\citenum{gaikwadEntanglementAssistedProbe2024}, and the design is posted on the open-source SQuADDS database \cite{shantoSQuADDSValidatedDesign2023}. We make use of a separate weakly-coupled microwave line to drive the qubit, which results in reasonably fast gate times ($80$~ns) with an average gate fidelity of 99.74\%. The device layout was created using the Qiskit Metal open-source package \cite{qiskit-metal} with the finite-element (FE) simulations taking place in the Ansys HFSS simulator to find the various parameters of the device, i.e., the qubit and cavity frequencies ($\omega_\mathrm{q}$ and $\omega_\mathrm{c}$), as well as the electric field distribution across the device of interest. The FE simulation results are then analyzed using the pyEPR package \cite{pyepr} to extract the qubit anharmonicity ($\eta$) and the qubit-cavity dispersive shift ($\chi_\mathrm{qc}$). Lastly, the resonator linewidth ($\kappa$) is extracted using the HFSS driven-modal scattering simulations. \cref{tab:pars} shows the measured parameters of the device used in this paper.

\begin{table} [h!]
\begin{center}
\begin{tabular}{| c | c | c | c |} 
\hline 
 \thead{$\omega_\mathrm{q}/2\pi$} & \thead{$|\eta|/2\pi$} & \thead{$\chi_\mathrm{qc}/2\pi$} & \thead{$\omega_\mathrm{c}/2\pi$}\\ 
\hline
  4.37 GHz & 215 MHz & 230 kHz & 6.95 GHz \\ 
 \hline
\end{tabular}
\begin{tabular}{| c | c | c |} 
\hline
\ \thead{$\kappa/2\pi$} & \thead{$T_1$} &\thead{$T_2^*$}  \\ 
\hline
  150 kHz & 20 $\mu\mathrm{s}$ & 32 $\mu\mathrm{s}$\\ 
 \hline
\end{tabular}
\end{center}
\caption{Measured parameters of the device used in the experiment.}
\label{tab:pars}
\end{table}
    
The device is fabricated and packaged in a copper box. We then mount and thermalize the box in the lowest stage of a dilution refrigerator (DR), which has a base temperature of $\sim 15 \mathrm{mK}$, surrounded by an additional copper shielding as well as a Cryoperm to further isolate the chip from the external infrared radiation and magnetic fields. The readout and drive lines each have 50 and 60 dB of attenuation to reach the single-photon power regime on the chip. Moreover, we use an 8 GHz low-pass filter on each line to protect the device from high-frequency noise, as well as high-frequency Eccosorb filters mounted close to the device in the DR to repel the infrared radiation from the higher stages of the DR. The readout line is protected by using four cryogenic microwave circulators, with the signal being amplified using both a traveling wave parametric amplifier (TWPA) at the mixing chamber and a high-electron-mobility transistor (HEMT) amplifier at the 4K stage to achieve single-shot readout.

\section{Conventional single-qubit gate calibration} 
\label{app:calib}

In our experiments, the $H_x$ Hamiltonian [\cref{eq:Hsys}] is fully calibrated and the $H_y$ Hamiltonian is then obtained simply by applying a phase shift via an arbitrary waveform generator (AWG), which is equivalent to $H_y = R_z(-\frac{\pi}{2}) H_x R_z(\frac{\pi}{2})$, and leaving $\epserr$ and $\delerr$ unchanged.

We use a cosine envelope pulse shape for all the experiments presented in this work. All our pulses are $80$~ns long with an additional padding of $4$~ns on both sides, i.e., $t_g=88$~ns. We perform the following steps to calibrate the single-qubit gates used in the experiments:

\begin{enumerate}
    \item Find the resonance frequencies of the cavity and the qubit using conventional spectroscopy techniques.
    \item Look for a monotone signal sent at the frequency of the qubit (cavity) at the output spectrum of the microwave mixers and try to minimize the leakage and the mirror signals caused by the mixer's electrical offset and imbalance.
    \item Perform simple Rabi experiments to find the right gate length for a specific gate amplitude. We chose a gate length of $80$~ns and a cosine-shaped gate envelope for the experiments reported here. Note that there is a padding of $8$~ns between consecutive pulses, and hence the effective pulse interval is $88$~ns. 
    \item Characterize the relaxation and decoherence rates of the device and use the Ramsey experiments to fine-tune the qubit's frequency.
    \item Optimize the readout fidelity by using the previously calibrated $R_x(\pi)$ pulses and varying the readout tone's frequency and amplitude.
    \item Apply consecutive $R_x(\pi)$ and $R_x(\pi/2)$ pulses to the system initialized in the ground state, such that after each measurement the qubit ideally returns to its ground state, i.e., $4N \times R_x(\pi/2)$ and $2N \times R_x(\pi)$ pulses with $N\in[1,\dots,20]$. Using this method we can observe the ground state fidelity of the qubit and pick an amplitude that pins the qubit to its ground state without exhibiting any oscillations. As described above, calibrating the pulses along the $x$-axis automatically calibrates them along the $y$-axis as well.   
    \item Optimize the DRAG weighting parameter ($\alpha$) by applying multiple pairs of $R_x(\pi) R_x(-\pi)$ pulses to the system initialized in its ground state \cite{Chen2016}, observe the ground state fidelity, and choose the parameter values for which the qubit remains in its ground state.
\end{enumerate}

\newpage
\begin{widetext}
\section{$\zeta$'s components}
\label{app:zeta}

The values of $\bz$'s components [defined in \cref{eq:29}], along with their first order approximations ($\delerr, \gamma, \epserr \ll \eps = \pi/t_g$) are noted below:

\begin{equation}
\begin{split}
\zeta_x &= \frac{-4 \eta e^{-2\gamma t_g}\delerr \epstot}{ \omega'^4 (\omega'^2 + \gamma^2)}\left[
\omega'^2 \left(
\left(\gamma \sin{\frac{\omega' t_g}{2}} +  \frac{(1-e^{\gamma t_g})}{2}\omega'\cos{\frac{\omega' t_g}{2}} \right)^2
+ \frac{\omega'^2}{4} \left( (1-e^{\gamma t_g})\sin{\frac{\omega' t_g}{2}}\right)^2
\right) \right.
\\ \notag
&\qquad - \left. \delerr^2 (1-\cos{(\omega' t_g)}) \left(
\left(\gamma \sin{\frac{\omega' t_g}{2}} +  \frac{1}{2}\omega' \cos{\frac{\omega' t_g}{2}} \right)^2
+ \frac{\omega'^2}{4} \left( 2(1-e^{\gamma t_g})-\cos^2{\frac{\omega' t_g}{2}}\right)
\right)
\right]
\\ \notag
&
\approx 0
\\
\zeta_y &= \frac{\eta e^{-2\gamma t_g }\omega' \epstot}{\omega'^4 (\omega'^2 + \gamma^2)} \left[
-4\delerr^2 \sin{(\omega' t_g)} 
\left(
\left(\gamma \sin{\frac{\omega' t_g}{2}} +  \frac{1}{2}\omega' \cos{\frac{\omega' t_g}{2}} \right)^2
+ \frac{\omega'^2}{4} \left( 2(1-e^{\gamma t_g})-\cos^2{\frac{\omega' t_g}{2}}\right)
\right) 
\right.
\\ \notag
&\qquad + \left. \gamma\omega'^3
\left(
\left(e^{\gamma t_g} - \cos{(\omega' t_g)} \right)^2
+ \sin^2{(\omega' t_g)}\right)
\right]
\\ \notag
&\approx (1+e^{-\gamma t_g})^2\frac{\gamma}{\eps}
\\
\zeta_z &= \frac{\eta e^{-2\gamma t_g}}{ \omega'^4 (\omega'^2 + \gamma^2)}  \left[
8\delerr^2 \epstot^2 \sin^2{\frac{\omega' t_g}{2}} 
\left(
 \left(\gamma \sin{\frac{\omega' t_g}{2}}  +  \frac{1}{2}\omega' \cos{\frac{\omega' t_g}{2}} \right)^2
- \frac{\omega'^2}{4} \cos^2{\frac{\omega' t_g}{2}}
\right) 
\right.
\\ \notag
&\qquad + \left. e^{\gamma t_g} \gamma\omega'^3
\left(
e^{\gamma t_g} \gamma\omega' + 2 \epstot^2 \sin{(\omega' t_g)}
\right)
+ \delerr^2 \omega'^2 (e^{\gamma t_g} - 1)
\left(
e^{\gamma t_g}\omega'^2 + (2\cos{(\omega' t_g)}-1) \epstot^2 + \delerr^2
\right)
\right]
\\ \notag
&\approx 0
\end{split}
\label{eq:zeta-values}
\end{equation}
\end{widetext}


\providecommand{\latin}[1]{#1}
\makeatletter
\providecommand{\doi}
  {\begingroup\let\do\@makeother\dospecials
  \catcode`\{=1 \catcode`\}=2 \doi@aux}
\providecommand{\doi@aux}[1]{\endgroup\texttt{#1}}
\makeatother
\providecommand*\mcitethebibliography{\thebibliography}
\csname @ifundefined\endcsname{endmcitethebibliography}
  {\let\endmcitethebibliography\endthebibliography}{}

\end{document}